\documentclass[conference]{IEEEtran}

\usepackage{xspace}
\newcommand{\sys}{\textsc{EvoPoC}\xspace}
\usepackage{threeparttable}
\usepackage{listings}
\usepackage{xcolor}
\usepackage{colortbl}
\usepackage[many]{tcolorbox}
\usepackage{booktabs}
\usepackage{multirow}
\usepackage{pifont}
\usepackage{multirow}
\usepackage{tabularx}
\usepackage{pifont}
\usepackage{subcaption} 
\usepackage{enumitem}
\usepackage{adjustbox}
\usepackage{amsmath}
\usepackage[linesnumbered,ruled,vlined]{algorithm2e}
\usepackage{hyperref}
\usepackage{amssymb}
\usepackage{placeins}

\definecolor{mygreen}{RGB}{0,205,0}

\lstdefinelanguage{Solidity}{
	keywords=[1]{receive, anonymous, assembly, assert, break, call, callcode, case, catch, class, constant, continue, constructor, contract, debugger, default, delegatecall, delete, do, else, emit, event, experimental, export, external, false, finally, for, function, gas, if, implements, import, in, indexed, instanceof, interface, internal, is, length, library, log0, log1, log2, log3, log4, memory, modifier, new, payable, pragma, private, protected, public, pure, push, require, return, returns, revert, selfdestruct, send, solidity, storage, struct, suicide, super, switch, then, this, throw, transfer, true, try, typeof, using, value, view, while, with, addmod, ecrecover, keccak256, mulmod, ripemd160, sha256, sha3}, % generic keywords including crypto operations
	keywordstyle=[1]\color{blue}\bfseries,
	keywords=[2]{address, bool, byte, bytes, bytes1, bytes2, bytes3, bytes4, bytes5, bytes6, bytes7, bytes8, bytes9, bytes10, bytes11, bytes12, bytes13, bytes14, bytes15, bytes16, bytes17, bytes18, bytes19, bytes20, bytes21, bytes22, bytes23, bytes24, bytes25, bytes26, bytes27, bytes28, bytes29, bytes30, bytes31, bytes32, enum, int, int8, int16, int24, int32, int40, int48, int56, int64, int72, int80, int88, int96, int104, int112, int120, int128, int136, int144, int152, int160, int168, int176, int184, int192, int200, int208, int216, int224, int232, int240, int248, int256, mapping, string, uint, uint8, uint16, uint24, uint32, uint40, uint48, uint56, uint64, uint72, uint80, uint88, uint96, uint104, uint112, uint120, uint128, uint136, uint144, uint152, uint160, uint168, uint176, uint184, uint192, uint200, uint208, uint216, uint224, uint232, uint240, uint248, uint256, var, void, ether, finney, szabo, wei, days, hours, minutes, seconds, weeks, years},	% types; money and time units
	keywordstyle=[2]\color{teal}\bfseries,
	keywords=[3]{block, blockhash, coinbase, difficulty, gaslimit, number, timestamp, msg, data, gas, sender, sig, value, now, tx, gasprice, origin},	% environment variables
	keywordstyle=[3]\color{violet}\bfseries,
	identifierstyle=\color{black},
	sensitive=true,
	comment=[l]{//},
	morecomment=[s]{/*}{*/},
	commentstyle=\color{gray}\ttfamily,
	stringstyle=\color{red}\ttfamily,
	morestring=[b]',
	morestring=[b]"
}

\ifCLASSINFOpdf
\else
\fi
\begin{document}

\author{
    \IEEEauthorblockN{
        Ruichao Liang\IEEEauthorrefmark{1},
        Jing Chen\IEEEauthorrefmark{2},
        Xianglong Li\IEEEauthorrefmark{2},
        Huangpeng Gu\IEEEauthorrefmark{2},
        Yebo Feng\IEEEauthorrefmark{1},
        Yue Xue\IEEEauthorrefmark{3},
        Cong Wu\IEEEauthorrefmark{2},
        Yang Liu\IEEEauthorrefmark{1}
    }
    \IEEEauthorblockA{
        \IEEEauthorrefmark{1}School of Computer Science and Engineering, Nanyang Technological University, Singapore\\
        \IEEEauthorrefmark{2}School of Cyber Science and Engineering, Wuhan University, Wuhan, China\\
        \IEEEauthorrefmark{3}MetaTrust Labs, Singapore
    }
}

\title{\sys: Automated Exploit Synthesis for DeFi Smart Contracts via Hierarchical Knowledge Graphs}

\IEEEoverridecommandlockouts
\makeatletter\def\@IEEEpubidpullup{6.5\baselineskip}\makeatother
\IEEEpubid{\parbox{\columnwidth}{
		Network and Distributed System Security (NDSS) Symposium 2026\\
		23 - 27 February 2026 , San Diego, CA, USA\\
		ISBN 979-8-9919276-8-0\\  
		https://dx.doi.org/10.14722/ndss.2026.[23$|$24]xxxx\\
		www.ndss-symposium.org
}
\hspace{\columnsep}\makebox[\columnwidth]{}}

\maketitle

\begin{abstract}

Smart contract vulnerabilities in Decentralized Finance (DeFi) caused over billions of dollars losses every year, yet the security community faces a critical bottleneck: identifying a vulnerability is not the same as proving it is exploitable.
Without an executable proof-of-concept (PoC) exploit, reported findings remain theoretical, leaving auditors unable to prioritize remediation.
Manual PoC construction is prohibitively labor-intensive, leaving most disclosed vulnerabilities unverified and protocols exposed long before mitigation is applied. 
While recent LLM-based approaches show promise, naive prompting falls short of producing reliable PoCs.

In this paper, we propose \sys, a knowledge-driven agentic system for end-to-end contract vulnerability detection and exploit synthesis.
Our core insight is that exploit synthesis is not a code generation task but a \emph{structured reasoning problem} that requires grounded knowledge of protocol semantics, failure root cause, and exploit primitives.
\sys organizes this knowledge into a \emph{Hierarchical Knowledge Graph} (HKG) that serves as structured memory for LLM-guided multi-hop reasoning, enabling the model to compose exploit strategies from reusable, semantically grounded primitives rather than unconstrained token generation.
To validate exploit feasibility beyond code synthesis, \sys employs a two-stage validation framework that checks exploit-path reachability via SMT solving and profit realizability via asset-level state simulation, ensuring generated PoCs satisfy both logical and economic viability constraints.
Evaluated on 88 real-world DeFi attacks and 72 audited projects (2,573 contracts), \sys achieves 98\% recall and 0.9 F1-score in detection, and a 96.6\% exploit success rate (ESR), reproducing 85 historical exploits and recovering over \$116.2M revenue. 
\sys outperforms SOTA fuzzers (\textsc{Verite}, \textsc{ItyFuzz}) by up to $5\times$ in ESR and $300\times$ in recoverable value, and the LLM-based exploit generator \textsc{A1} by $2\times$ and $8.5\times$ respectively.
In bug bounty evaluation, \sys identified 16 confirmed 0-day vulnerabilities, helping secure over \$70.6M and earning \$2,900 in bounties.

\end{abstract}

\IEEEpeerreviewmaketitle

\section{Introduction}

DeFi protocols have become one of the most lucrative targets for adversarial exploitation in the history of software security, with an estimated \$3.6 billion lost to smart contract flaws in 2025 alone~\cite{Crystal}.
Numerous automated techniques, including static analysis~\cite{Slither,ghaleb2023achecker,qin2025enhancing,xie2024defort}, fuzzing~\cite{shou2023ityfuzz,10.1145/3715714}, and formal verification~\cite{stephens2021smartpulse,bose2022sailfish,DBLP:conf/ndss/0012XW00S025,zhang2024towards}, have been proposed to identify potential vulnerabilities. 
However, in practice, these tools produce large volumes of unverified alerts, and security teams must manually assess each finding to determine whether it represents an actual vulnerability.
This manual triage process is slow, expensive, and error-prone.

Large language models (LLMs) offer a promising path toward automating this confirmation step. 
However, existing LLM-based approaches~\cite{10.1145/3597503.3639117,11019687,jin2025llmbscvmllmbasedblockchainsmart,10.1109/ICSE55347.2025.00027,david2023needmanualsmartcontract} focus narrowly on vulnerability \emph{detection} and lack the capability for end-to-end exploit synthesis. 
Gervais et al.~\cite{gervais2025aiagentsmartcontract} take a step toward exploit generation via tool-calling, but their system relies predominantly on the LLM's intrinsic capability, making it prone to hallucination, unstable reasoning, and limited robustness and reproducibility in complex scenarios.
The fundamental problem is that exploit generation in DeFi is not a code completion task: it is a multi-step reasoning problem that requires precise knowledge of protocol semantics, economic models, and end-to-end executability.

To fill the gap, we propose \sys, a knowledge-driven agentic system for automated end-to-end smart contract exploit synthesis.
Rather than viewing exploit generation as a direct extension of vulnerability detection or a one-shot code generation, we formulates it as a structured reasoning problem that retrieves semantic context from grounded domain knowledge, synthesizes exploit primitives through multi-hop reasoning, and verifies their logical and economic feasibility.
Specifically, we make several innovations to address three core challenges:

\textbf{i) The semantic gap between vulnerability detection and exploit synthesis.}
Since LLM training knowledge is implicitly distributed across model parameters, its reasoning capability largely manifests as probabilistic navigation in high-dimensional semantic spaces~\cite{10.5555/3666122.3669203}. 
In complex DeFi protocols, however, vulnerability detection and exploitation are separated by a substantial semantic distance:
the former focuses on localized anomalies within contract implementations~\cite{10.1145/3728878,10.1145/3728924,10.1145/3597503.3639117}, whereas the latter requires synthesizing multi-contract interactions, protocol states, and economic constraints to construct executable attack strategies~\cite{FORAY}. 
This mismatch significantly amplifies reasoning errors as the model attempts to navigate sparse semantic connections, which is a phenomenon known as probability diffusion~\cite{10.5555/3495724.3496298}.
To bridge this gap, we define an ontology schema that organizes DeFi domain knowledge into a structured, explicit representation called Hierarchical Knowledge Graph (HKG). 
By structuring knowledge along three dimensions: contract semantics, failure modes, and exploit primitives, the HKG provides reusable symbolic anchors that enable multi-hop reasoning and guide LLMs from vulnerability detection through root cause analysis to composition of exploit primitives.

\textbf{ii) Hallucination amplification in long-horizon reasoning chains.}
The effectiveness of LLM-based approaches~\cite{gervais2025aiagentsmartcontract,10.1145/3597503.3639117,11019687,jin2025llmbscvmllmbasedblockchainsmart,10.1109/ICSE55347.2025.00027} is often limited by hallucinations when reasoning over multi-step chains: a single fabricated primitive or misinterpreted protocol behavior compounds across subsequent steps, collapsing the entire exploit logic.
The challenge intensifies as novel DeFi attack patterns continually outpace model training data.
To address this, we develop an evolving agentic memory mechanism built upon the HKG, consisting of long-term memory (LTM) and working memory (WM). 
The system continuously updates LTM by extracting structured and reusable knowledge from the latest audit reports and exploits under the predefined HKG ontology, and dynamically instantiates task-specific working memory during contract analysis.
This mechanism assists in establishing a coherent and grounded reasoning chain, significantly reducing hallucinations with up-to-date domain expertise.

\textbf{iii) Lack of practical exploit validity guarantees.}
LLM-generated PoCs are frequently syntactically plausible but semantically invalid: they reference non-existent execution paths, fail to satisfy state prerequisites, or produce no net profit under realistic execution conditions. 
To address this, \sys incorporates a two-stage validation framework that verifies \emph{exploit-path reachability} via SMT constraint checking and \emph{profit realizability} via asset-level state simulation, filtering infeasible candidates before Foundry execution and providing informative feedback for iterative refinement.

\sys’s long-term memory is bootstrapped from audit reports and real-world exploit analyses across multiple sources~\cite{openzeppelin,chainalysis,certikt,medium,x} through automated knowledge extraction, and can be continuously updated.
We evaluated \sys on 88 real-world DeFi attack incidents and 72 audited projects comprising 2,573 contracts\footnote{Evaluation datasets were not used in constructing the LTM of \sys.}, comparing it with state-of-the-art (SOTA) tools, including LLM-based scanner \textsc{GPTScan}~\cite{10.1145/3597503.3639117}, the fuzzers \textsc{ItyFuzz}~\cite{shou2023ityfuzz} and \textsc{Verite}~\cite{VERITE}, and the LLM-based exploit generation tool \textsc{A1}~\cite{gervais2025aiagentsmartcontract}.
\sys outperforms \textsc{GPTScan} by achieving a recall of 98\% and an F1-score of 90\% in vulnerability detection.
Its generated PoCs successfully reproduce 85 historical real-world exploits, yielding a total profit of \$116.2 million.
\sys attains about 5$\times$ higher exploit success than \textsc{ItyFuzz}, and 2$\times$ higher than \textsc{Verite} and \textsc{A1}, while achieving over 300$\times$, 2$\times$, and 8.5$\times$ higher revenue, respectively.
In addition, \sys discovered 21 0-day vulnerabilities, 16 of which have been acknowledged or patched, helping secure \$70.6 million and earning \$2,900 in bug bounties.

\textbf{Contributions.} This paper makes following contributions.
\begin{itemize}[leftmargin=4mm, itemindent=0mm]
    \item We demystify three root causes that make automated exploit synthesis fundamentally harder than vulnerability detection in DeFi: 
    \ding{172}~the semantic distance between localized flaw identification and complex PoC composition; 
    \ding{173}~LLM susceptibility to hallucination and knowledge staleness during long-horizon security reasoning; and
    \ding{174}~the absence of executable validity guarantees for LLM-generated PoCs in realistic execution environments.

    \item We make three innovations to address these gaps, including 
    \ding{172} a hierarchical knowledge graph that encoding DeFi security knowledge as structured reasoning anchors to bridge vulnerability detection and exploit synthesis;
    \ding{173} an evolving agentic memory mechanism to ground LLM reasoning and reduce hallucinations with up-to-date knowledge distilled from real-world audits and exploits; and
    \ding{174} a two-stage feasibility validation framework combining SMT-based reachability checking with profit realizability simulation.

    \item We build and evaluate \sys, an agentic system for end-to-end vulnerability identification and exploit synthesis.
    Experiments show that it outperforms SOTA vulnerability scanner in detection accuracy, surpasses leading fuzzers and PoC generators in exploit effectiveness, reproduces 85 real-world exploits extracting over \$116.2M, and discovers 16 0-day vulnerabilities yielding \$2,900 in bug bounties.
  \end{itemize}

\section{Background}

\subsection{Smart Contract and Decentralized Finance}
Smart contracts manage digital assets on the blockchain via predefined logic~\cite{10.1145/3558535.3559780}, underpinning the DeFi ecosystem of lending, trading, and asset management~\cite{john2023smart,leon2025data}.
Logic vulnerabilities are frequently exploited, causing substantial losses~\cite{10179435}.
Validating such vulnerabilities requires a proof-of-concept (PoC) exploit, yet PoC construction demands deep DeFi expertise and compositional reasoning, making it costly and hard to scale.

\subsection{LLM-based Vulnerability Detection}
LLMs have been applied to vulnerability detection via fine-tuning, chain-of-thought reasoning, and integration with static analysis or symbolic execution~\cite{10988968,lin2025large,10.1145/3597503.3639117,11019687,jin2025llmbscvmllmbasedblockchainsmart}.
However, these methods identify potential vulnerabilities but struggle to assess exploitability or generate valid PoCs, limiting their practical utility.

\section{Threat Model}

\textbf{Adversary model.}
We consider an adversary seeking financial gain by exploiting vulnerabilities in publicly deployed DeFi smart contracts.
The adversary has read access to on-chain bytecode, contract ABIs, historical transactions, and public disclosures, and aims to construct transaction sequences that triggers a target vulnerability and extracts unauthorized profit. 
This model captures the real-world attacker profile observed in historical DeFi incidents.

\textbf{Defender model.}
\sys is designed as a proactive defensive security tool that enables authorized security practitioners to stay ahead of adversaries through an end-to-end pipeline covering both vulnerability identification and exploit confirmation.
Intended users include security auditors, bug bounty hunters, and protocol teams conducting internal reviews. 
In all cases, practitioners operate on contracts they are authorized to test, and exploits execute only in controlled local fork environments.

\begin{lstlisting}[language = Solidity, caption=Fee-on-transfer token with delayed distribution causing reserve imbalance and price manipulation.,label=lst 1, float =t!, label = lst:motivating_example]
function buyTokenAndFees(address from, address to, uint256 amount) internal {
    uint256 burnAmount = amount.mul(3).div(100); 
    uint256 otherAmount = amount.mul(1).div(100); 
    uint256 feeAmount = amount.mul(10).div(100);  
    amount = amount.sub(feeAmount);    
    swapFeeTotal = swapFeeTotal.add(otherAmount);
    super._burn(from, burnAmount);
    super._transfer(from, to, amount);}

function distributeFee() public {
    uint256 mokeyFeeTotal = swapFeeTotal.mul(2);
    super._transfer(uniswapV2Pair, monkeyWallet, mokeyFeeTotal);
    super._transfer(uniswapV2Pair, birdWallet, swapFeeTotal);
    super._transfer(uniswapV2Pair, foundationWallet, swapFeeTotal);
    super._transfer(uniswapV2Pair, technologyWallet, swapFeeTotal);
    super._transfer(uniswapV2Pair, marketingWallet, swapFeeTotal);
    swapFeeTotal = 0;}
  \end{lstlisting}

\section{Motivating Example}
Despite the capabilities of LLMs in understanding smart contract code, a substantial gap remains between identifying potential vulnerabilities and generating practical PoCs.
Here, we show how structured and reusable knowledge enables compositional reasoning in LLMs and bridge this gap.

\begin{figure}[t!]
    \centering
    \begin{tcolorbox}[title=\footnotesize{CoT-based-only Prompt Template}, before upper=\scriptsize]
        \vspace{-3pt}
        \textbf{System:}
        You are a professional smart contract auditor ...

        \textbf{Step 1:}
        Review the contract code \textbf{[\%CODE\%]} and and identify any logic flaws or security issues that could lead to unexpected or exploitable behavior. 
        Report only the most confident and impactful finding.
        
        \textbf{Step 2:}
        Based on the identified issue, analyze its root cause and evaluate whether it can be exploited for financial gain or manipulation in a DEX or liquidity pool setting. 
        If so, outline a high-level attack strategy.

        \textbf{Step 3:}
        Based on the strategy, generate a concise proof-of-concept in Solidity that demonstrates the sequence of interactions an attacker would use.
        Use the environment assumptions \textbf{[\%ENV\%]} during PoC generation.
        \vspace{-3pt}
    \end{tcolorbox}
    \vspace{-10pt}
    \caption{Step-wise prompt for guiding LLMs to identify and exploit the vulnerability in the motivating example.}
    \label{fig:cot_prompt}
\end{figure}

\subsection{Detection-to-Exploitation Gap}

Listing~\ref{lst:motivating_example} shows a token contract snippet from a real-world price manipulation incident.
The token implements a fee-on-transfer mechanism that deducts 10\% per transfer, burning 3\% and redistributing the remaining 7\% to designated wallets.
In practice, the 7\% is temporarily held in sender's account and only released via \texttt{distributeFee}, creating a mismatch between the sender's actual balance and deserved balance.

While detecting this anomaly in \texttt{buyTokenAndFees} is straightforward for LLMs, exploiting it is more complex.
An attacker must first locate an AMM pair, obtain capital via a flash loan, and repeatedly \texttt{transfer} tokens into the pair to accumulate the retained 7\% fees inside, and then invoke \texttt{skim} to reclaim the transferred amount.
Finally, the attacker triggers \texttt{distributeFee} to induce a sharp reserve imbalance, and distort the AMM price to arbitrage profits.
Notably, although the vulnerability originates from the token’s fee-handling logic, the exploitation does not manifest locally. 
Instead, it unfolds within an external AMM pair through a sequence of coordinated interactions across contracts.
The attack requires reasoning over fee-on-transfer semantics, AMM pricing, flash loan, and adversarial state manipulation.
This separation between localized vulnerability detection and cross-contract, stateful exploitation illustrates a substantial semantic gap that is difficult to bridge through LLM reasoning alone.

We further use a CoT-based prompt (Figure~\ref{fig:cot_prompt}) and GPT-5 to conduct an evaluation on this example.
Table~\ref{tab:motivating} summarizes the model’s performance across successive reasoning stages.
While the model consistently recognizes the presence of the price manipulation vulnerability (5/5), its performance degrades sharply when deeper causal reasoning (3/5) and exploit construction (0/5) are required, and no executable PoC is produced (0/5).
This result provides concrete evidence of a substantial reasoning gap between vulnerability detection and practical exploitation.

\renewcommand{\arraystretch}{0.9}
\begin{table*}[t!]
    \centering
  \scriptsize
  \setlength{\tabcolsep}{10pt}
\caption{Evaluation of LLM reasoning progression on the motivating example. C-B (CoT-based-only) and K-G (Knowledge-guided) were each tested in 5 runs with up to 5 iterations. }
\label{tab:motivating}
\begin{tabular}{@{}cccccl@{}}
\toprule
\multirow{2}{*}{\textbf{Reasoning Stage}} &
  \multicolumn{2}{c}{\textbf{Result}} &
  \multicolumn{2}{c}{\textbf{Iteration}} &
  \multicolumn{1}{c}{\multirow{2}{*}{\textbf{Notes}}} \\ \cmidrule(lr){2-3} \cmidrule(lr){4-5}
 &
  C-B &
  K-G &
  C-B &
  K-G &
  \multicolumn{1}{c}{} \\ \midrule
\begin{tabular}[c]{@{}c@{}}Vulnerability Identification \end{tabular} &
  5/5 &
  \textbf{5/5} &
  1.8 &
  \textbf{1} &
  \begin{tabular}[c]{@{}l@{}}Correctly recognizes the presence and category of the vulnerability.\end{tabular} \\ \midrule
\begin{tabular}[c]{@{}c@{}}Root Cause Analysis \end{tabular} &
  3/5 &
  \textbf{5/5} &
  4 &
  \textbf{1.4} &
  \begin{tabular}[c]{@{}l@{}}Correctly analyze the root cause of the vulnerability.\end{tabular} \\ \midrule
Exploit Strategy Synthesis &
  0/5 &
  \textbf{5/5} &
  5 &
  \textbf{2} &
  \begin{tabular}[c]{@{}l@{}}Synthesizes the attack strategy spanning flash loans, token transfers, \\and AMM interactions.\end{tabular} \\ \midrule
Executable PoC Generation &
  0/5 &
  \textbf{4/5} &
  5 &
  \textbf{2} &
  \begin{tabular}[c]{@{}l@{}}Produces an executable PoC that succeeds under realistic environment.\end{tabular} \\ \bottomrule
\end{tabular}
\end{table*}

\renewcommand{\arraystretch}{1}

\begin{figure}[t!]
    \centering
    \begin{tcolorbox}[title=\footnotesize{Structured Knowledge-guided Prompt Template}, before upper=\scriptsize]
      \vspace{-3pt}
        \textbf{System:}
        You are a professional smart contract auditor ...

        \textbf{Step 1:}
        Review the protocol context nodes \textbf{[\%PROTOCOL\%]} and contract code \textbf{[\%CODE\%]}, select only the nodes that are semantically consistent with the contract’s logic and interaction patterns.
        
        \textbf{Step 2:}
        Based on the selected protocol nodes and contract code, determine whether it exhibits vulnerabilities described in the failure mode nodes \textbf{[\%FAIL\%]}. 
        Retain those that genuinely match and link them explicitly with the corresponding protocol nodes.

        \textbf{Step 3:}
        Based on the result, select exploit primitives from \textbf{[\%EXP\%]} that can trigger the vulnerability. 
        Retain those that genuinely match and link them with the corresponding root cause nodes.
        Compose these primitives into a coherent high-level attack strategy.

        \textbf{Step 4:}
        Based on the strategy, generate a concise proof-of-concept in Solidity that demonstrates the sequence of interactions an attacker would use.
        Use the environment assumptions \textbf{[\%ENV\%]} during PoC generation.
        \vspace{-3pt}
    \end{tcolorbox}
        \vspace{-10pt}
    \caption{Structured knowledge as reasoning primitives for vulnerability detection and exploitation.}
    \label{fig:knowledge_prompt}
\end{figure}

\begin{figure}[t!]
  \centering
  \includegraphics[width=3.2in]{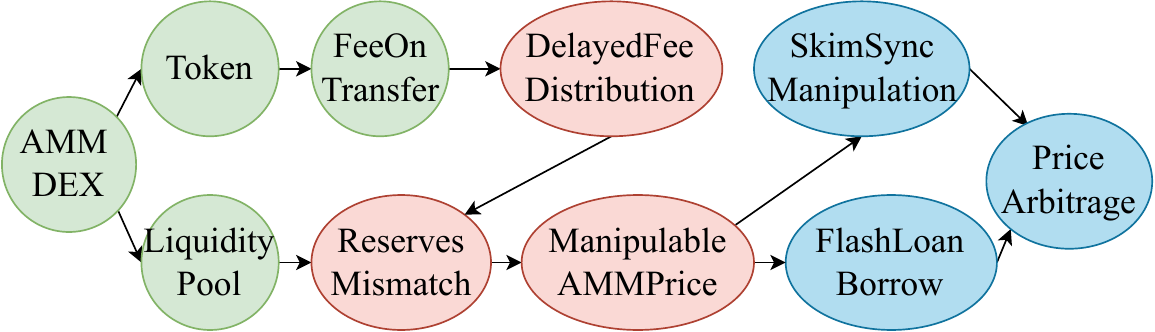}
  \caption{Reasoning chains from structured knowledge nodes to guide exploit synthesis. Green, red, and blue nodes represent [\%PROTOCOL\%], [\%FAIL\%], and [\%EXP\%] types.}  
  \label{fig:motivating}
\end{figure}

\begin{figure*}[t!]
  \centering
  \includegraphics[width=7in]{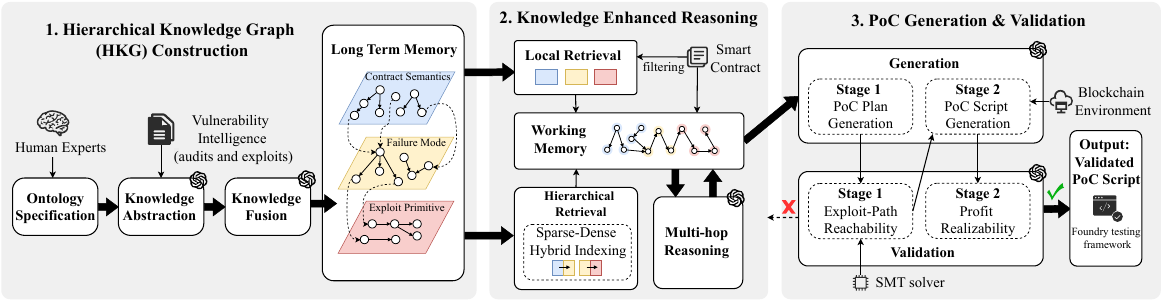}
  \caption{Overview of \sys.}
  \label{fig:overview}
\end{figure*}

\subsection{Structured Knowledge as Semantic Link}

We augment the LLM with structured domain knowledge relevant to this example.
Specifically, we abstract DeFi protocol context, token–AMM interactions, typical fee-on-transfer flaw root causes, and common state manipulation and arbitrage strategies into structured knowledge entries.
As shown in Figure~\ref{fig:knowledge_prompt}, these entries are exposed to the LLM as candidate reasoning primitives during inference, allowing the model to select and compose them rather than being directly guided toward the ground-truth exploit.
As illustrated in Figure~\ref{fig:motivating}, the LLM selectively extracts relevant knowledge nodes and composes them into explicit reasoning chains. 
These chains operationalize structured knowledge as intermediate reasoning steps rather than implicit heuristics. 
As shown in Table~\ref{lst:motivating_example}, this knowledge-guided setting (K‑G) consistently outperforms the baseline (C‑B), achieving higher success rates across all reasoning stages while requiring fewer iterations.

The result demonstrates that structured and reusable knowledge can provide semantic link that enable LLMs to perform compositional inference and supply the necessary reasoning primitives, effectively bridging the gap between vulnerability detection and exploit construction.

\section{Methodology}

\subsection{Overall Workflow}
As shown in Figure~\ref{fig:overview}, \sys consists of three phases: hierarchical knowledge graph (HKG) construction, knowledge-enhanced multi-hop reasoning, as well as two-stage PoC generation and validation.

In the first phase, \sys constructs HKG in three steps.
i) For ontology specification, ontology schema is defined to specify a three-layer structure, including node and edge types and their connection constraints, establishing an inductive schema for knowledge construction.
ii) In knowledge abstraction, guided by this ontology, the LLM extracts structured nodes and edges from real-world vulnerability intelligence such as audit reports and exploits, transforming unstructured text into reusable abstract knowledge.
iii) Finally, knowledge from multiple sources is merged by aligning nodes, removing redundancies, and resolving conflicts, producing a unified and consistent HKG that serves as long-term memory.

The second phase illustrates the knowledge-enhanced reasoning process of \sys.
Given a contract project, \sys first filters candidate functions to exclude non-Solidity files, test code, and trusted libraries. 
It then builds a task-specific working memory by incrementally retrieving relevant HKG nodes to improve LLM reasoning.
The process starts with local retrieval of a candidate node in the first HKG layer. 
Based on this node, the LLM performs multi-hop reasoning to identify additional relevant nodes, which are incorporated into the working memory to incrementally build the layer-specific subgraph.
After completing reasoning within a layer, hierarchical retrieval selects candidate nodes for the next layer, and the interleaved, memory-guided retrieval–reasoning process repeats, gradually constructing a coherent reasoning subgraph that serves as the working memory for subsequent PoC generation.

In the third phase, \sys interleaves PoC generation and validation.
Leveraging the reasoning subgraph and contract code in the working memory, the LLM first produces a high-level exploit plan, which is validated through path-level analysis to ensure reachability.
Based on the validated plan, a concrete exploit script is generated and further verified through state-based analysis by tracking account-level fund flows and state transitions to confirm the intended exploit effects. 
Only PoCs that pass both validation stages are retained for execution in an on-chain environment.

\begin{figure*}[t!]
  \centering
  \includegraphics[width=7in]{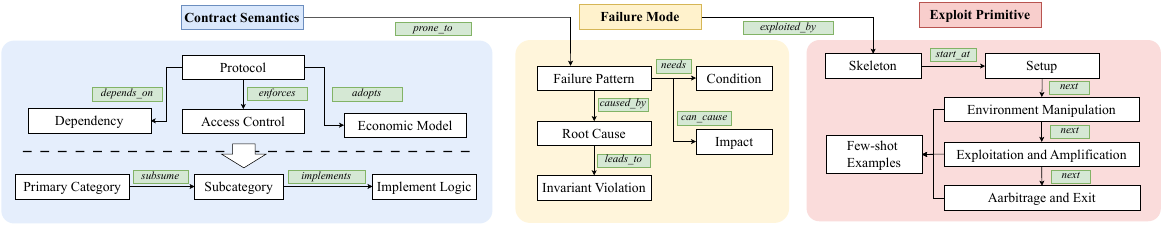}
  \caption{Overview of the hierarchical knowledge graph (HKG) ontology schema.}
  \label{fig:ontology}
\end{figure*}

\subsection{Hierarchical Knowledge Graph}
\label{sec:HKG}

Hierarchical knowledge graph serves as the core representation in \sys’s memory.
It abstracts concrete DeFi semantics and vulnerability intelligence into structured, reusable knowledge, which can be continuously refined through automated knowledge construction and fusion.

\textbf{Ontology schema}.
As shown in Figure~\ref{fig:ontology}, we model the HKG ontology as a typed directed graph $\mathcal{O} = \langle \mathcal{V}, \mathcal{E}, \mathcal{T} \rangle$,
where $\mathcal{V}$ denotes a set of nodes, $\mathcal{E}$ denotes a set of directed edges, and $\mathcal{T}$ defines node types, edge types, and their admissible connections.
Nodes are organized into three semantic layers: \emph{contract semantics}, \emph{failure mode}, and \emph{exploit primitive}, corresponding to progressively deeper stages of vulnerability understanding.

i) For the \textit{contract semantics layer}, we define $\mathcal{S} = \{\mathit{Prot}, \mathit{Acc}, \mathit{Eco}, \mathit{Dep}\}$ as the set of node types, corresponding to \emph{protocol}, \emph{access control}, \emph{economic model}, and \emph{dependency}, respectively.
Protocol node $p \in \mathit{Prot}$ serves as a semantic anchor, connected to other semantic nodes via typed relations:
\begin{equation}
  \small
\begin{aligned}
&\mathit{enforces} : \mathit{Prot} \rightarrow \mathit{Acc}, \quad
\mathit{adopts} : \mathit{Prot} \rightarrow \mathit{Eco}, \\
&\mathit{depends\_on} : \mathit{Prot} \rightarrow \mathit{Dep}.
\end{aligned}
\end{equation}
To support fine-grained abstraction, each semantic node $c \in \mathcal{C}$ is internally structured as $c = \langle c_p, c_s, l \rangle$,
where $c_p \in \mathit{PrimaryCategory}$, $c_s \in \mathit{SubCategory}$, and $l \in \mathit{ImplementLogic}$.
The hierarchical relations explicitly encode this decomposition:
\begin{equation}
  \small
\begin{aligned}
  &\mathit{subsume} : \mathit{PrimaryCategory} \rightarrow \mathit{SubCategory}, \\
  &\mathit{implements} : \mathit{SubCategory} \rightarrow \mathit{ImplementLogic}
\end{aligned}
\end{equation}

ii) For the \textit{failure mode layer}, we define $\mathcal{F} = \{\mathit{FP}, \mathit{Cond}, \mathit{Imp}, \mathit{RC}, \mathit{Inv}\}$ as the set of node types, corresponding to \emph{failure pattern}, \emph{condition}, \emph{impact}, \emph{root cause}, and \emph{invariant violation}, respectively.
Each failure pattern $fp \in \mathit{FP}$ is connected by typed relations:
\begin{equation}
  \small
\begin{aligned}
&\mathit{caused\_by} : \mathit{FP} \rightarrow \mathit{RC}, \quad
\mathit{needs} : \mathit{FP} \rightarrow \mathit{Cond}, \\
&\mathit{can\_cause} : \mathit{FP} \rightarrow \mathit{Imp}.
\end{aligned}
\end{equation}
Root cause $rc \in \mathit{RC}$ is connected to an invariant violation by:
\begin{equation}
  \small
\mathit{leads\_to} : \mathit{RC} \rightarrow \mathit{Inv}.
\end{equation}

iii) For the \textit{exploit primitive layer}, let $\mathcal{X} = \{\mathit{Skel}, \mathit{Prim}, \mathit{Ex}\}$ denote the set of node types in the exploit primitive layer.
Each exploit skeleton $sk \in \mathit{Skel}$ is connected to a sequence of exploit primitives: $\langle p_0, p_1, \dots, p_n \rangle$, where $\ p_i \in \mathit{Prim}$, via typed relations:
\begin{equation}
  \small
\mathit{start\_at} : \mathit{Skel} \rightarrow \mathit{Prim}, \quad
\mathit{next} : \mathit{Prim} \rightarrow \mathit{Prim}.
\end{equation}
Exploit primitive $p \in \mathit{Prim}$ is assigned a semantic role from:
\begin{equation}
  \small
\begin{aligned}
\{\,&\mathit{Setup},\ \mathit{EnvironmentManipulation}, \\
&\mathit{ExploitationAndAmplification},\ \mathit{ArbitrageAndExit}\,\}.
\end{aligned}
\end{equation}
Each skeleton is further linked to few-shot examples by:
\begin{equation}
  \small
\mathit{illustrated\_by} : \mathit{Skel} \rightarrow \mathit{Ex}.
\end{equation}

Cross-layer edges are introduced selectively to encode causal relevance rather than exhaustive linkage.
Specifically, a contract semantics node is linked to a failure pattern only if it contributes to the pattern's root cause, and a failure pattern is linked to an exploit skeleton only if the pattern is necessary for the exploit.
This design supports structured reasoning from high-level semantics to concrete exploit realizations.

\begin{figure*}[t!]
  \centering
  \includegraphics[width=6in]{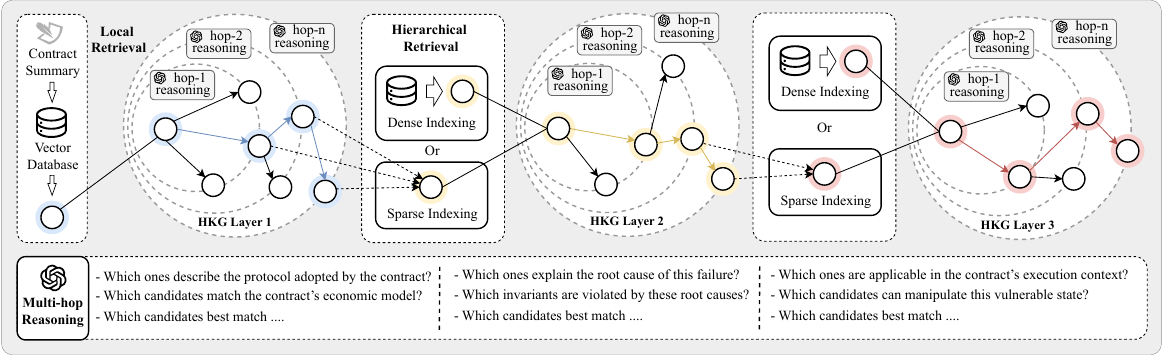}
  \caption{Multi-hop reasoning workflow based on agentic memory, where retrieved knowledge from the HKG (LTM) is organized into a task-specific subgraph as \sys's working memory (WM).}
  \label{fig:Multi-hop}
\end{figure*}

\textbf{HKG construction and evolution}.
Based on the ontology schema, we leverage LLMs to instantiate concrete nodes and edges in the HKG.
Input knowledge is collected from real-world vulnerability intelligence, including audit reports~\cite{openzeppelin,chainalysis}, attack incident disclosures~\cite{certikt,x}, and technical analysis blogs~\cite{medium}. 
A case is selected if it satisfies at least two of the following criteria: i) availability of the vulnerable contract source code, ii) detailed analysis of the vulnerability or attack event, and iii) explicit descriptions of exploitation steps or exploit script.
The construction process consists of knowledge abstraction and knowledge fusion. 

In the abstraction stage, given unstructured textual intelligence for a case, we prompt LLMs with chain-of-thought instructions to extract ontology-conformant nodes and relations.
The model is explicitly required to assign each extracted element a node type in $\mathcal{S}$, $\mathcal{F}$, or $\mathcal{X}$, a relation type in $\mathcal{E}$, and a concise semantic description.
For contract semantics, node descriptions include the intended functionality and potential defect-prone aspects.
For failure mode nodes, descriptions encode the failure pattern together with its semantic context and triggering conditions.
For exploit primitive nodes, descriptions summarize the exploit behavior, the exploited failure mode, and the resulting impact.
When certain ontology elements are missing from the input intelligence, the model may infer and complete them through contextual reasoning.
Each case is instantiated as an individual, case-level subgraph of $\mathcal{O}$.

In the fusion stage, case-level subgraphs are incrementally integrated into the global HKG.
For each newly generated node $v\in\mathcal{V}$, the LLM determines whether it corresponds to an existing node by jointly considering its node type, semantic descriptions, and neighboring relations.
Candidate matching proceeds in two stages: a \emph{type-filtered sparse lookup} first restricts the search to nodes within the same ontology sublayer, and a \emph{Faiss ANN search}~\cite{Faiss} over node-description embeddings then retrieves the top-$k$ semantic neighbors.
This design avoids pairwise comparison across the full graph, allowing fusion cost to scale sub-linearly with graph size.
Nodes judged semantically equivalent are merged with their incident edges.
Nodes sharing high-level intent but differing in scope or granularity are preserved as parallel variants.
Edges are merged only when both endpoints are aligned.
The same abstraction and fusion procedures incorporate newly collected vulnerability intelligence, allowing the graph to evolve continuously as new cases emerge.

The complete ontology schema, node type definitions, and HKG statistics are provided in Appendix~\ref{app:hkg}.

\subsection{Memory-Enhanced Multi-hop Reasoning}

The HKG constitutes the long-term memory (LTM) of \sys.
It is maintained as a graph that encodes ontology-conformant nodes and their relations, with node descriptions indexed by vector embeddings.
Given a target contract, relevant knowledge is selectively retrieved from the LTM and instantiated into the working memory (WM), enabling the LLM to perform agentic multi-hop reasoning.
This process produces a case-specific, holistic understanding of the contract, including its semantics, relevant failure modes, and corresponding exploit strategies.

As shown in Figure~\ref{fig:Multi-hop}, reasoning is initiated by generating a semantic summary of the target contract.
In the contract semantics layer, similarity-based retrieval identifies the most relevant protocol-level primary category nodes, which serve as initial reasoning seeds.
Starting from these seeds, the LLM incrementally expands a protocol-level subgraph via multi-hop traversal of graph relations, guided by the contract context.
At each hop, one-hop neighboring nodes are retrieved as candidates and evaluated under CoT reasoning to determine their relevance.
This iterative expansion completes local retrieval within the layer.
Cross-layer transitions are realized through hierarchical retrieval using a sparse-dense hybrid strategy.
To move from the contract semantics layer to the failure mode layer, sparse retrieval first checks whether multiple selected semantic nodes converge on the same failure pattern through existing cross-layer links.
If a candidate exceeds a predefined confidence threshold, it is selected as the next reasoning seed.
When sparse retrieval yields no candidate, dense retrieval is applied by performing similarity search over failure pattern nodes using semantic embeddings, from which the LLM selects the most plausible node conditioned on the contract context.
The same retrieval-and-selection procedure is applied when transitioning from the failure mode layer to the exploit primitive layer.

Within each layer, reasoning follows a breadth-first strategy.
At each step, candidate nodes are evaluated against layer-specific criteria.
In the contract semantics layer, the LLM assesses whether the contract adopts the protocol type, economic model, or access control pattern represented by a candidate node.
In the failure mode layer, it evaluates whether the contract behavior aligns with the vulnerability pattern captured by the candidate.
In the exploit primitive layer, it selects exploitation strategies that are consistent with the identified failure modes.
Reasoning within a layer terminates when no candidate nodes remain consistent with the contract context or when further expansion is deemed uninformative.

\begin{algorithm}[t]
\scriptsize
\caption{\small{Executable exploit synthesis with feasibility}}
\label{alg:exploit_synthesis}

\KwIn{Working memory $\mathcal{WM}$, execution environment $\mathcal{E}$}
\KwOut{Exploit script: $\textit{script}$ or \textbf{Failure}}
\SetFuncSty{textsc}

\SetKwProg{Fn}{Procedure}{:}{}
\SetKwFunction{ExploitSynthesis}{ExploitSynthesis}
\SetKwFunction{CheckPathReachability}{CheckPathReachability}
\SetKwFunction{GenerateExploitPlan}{GenerateExploitPlan}
\SetKwFunction{GenerateExploitScript}{GenerateExploitScript}
\SetKwFunction{ValidateProfitability}{ValidateProfitability}
\SetKwFunction{DeployToFoundry}{DeployToFoundry}
\SetKwFunction{SemanticTraversal}{SemanticTraversal}
\SetKwFunction{CollectPredicates}{CollectPredicates}
\SetKwFunction{SMTSatisfiable}{SMTSatisfiable}
\SetKwFunction{SimulateExecution}{SimulateExecution}

\Fn{\ExploitSynthesis{$\mathcal{WM}, \mathcal{E}$}}{
    $\mathcal{P} \gets$ \GenerateExploitPlan{$\mathcal{WM}$}\;
    $\textit{reachable} \gets$ \CheckPathReachability{$\mathcal{P}, \mathcal{E}$}\;

    \If{$\textit{reachable} = \textbf{true}$}{
        $\textit{script} \gets$ \GenerateExploitScript{$\mathcal{P}, \mathcal{E}$}\;
        \If{\ValidateProfitability{$\textit{script}, \mathcal{E}$} = \textbf{true}}{
            \DeployToFoundry{$\textit{script}$}\tcc*[r]{Forward for execution}
        }
        \lElse{
            \Return \textbf{Failure}\tcc*[f]{Not profitable}}
    }
    \lElse{
        \Return \textbf{Failure}\tcc*[f]{Path infeasible}
    }
}

\Fn{\CheckPathReachability{$\mathcal{P}, \mathcal{E}$}}{
    \ForEach{$t = \langle \mathcal{C}, f, \sigma, \mathcal{K} \rangle \in \mathcal{P}$}{
        $\pi \gets$ \SemanticTraversal{$\mathcal{C}, f, \mathcal{K}$}\;
        $\Phi \gets$ \CollectPredicates{$\pi$}\;
        \If{\textbf{not} \SMTSatisfiable{$\bigwedge \Phi$}}{
            \Return \textbf{false}\tcc*[f]{Unreachable sink}
        }
    }
    \Return \textbf{true}
}

\Fn{\ValidateProfitability{$\textit{script}, \mathcal{E}$}}{
    Initialize $\mathcal{S} = \langle \mathcal{B}_{init}, \Omega_{init} \rangle$\;
    \SimulateExecution{$\textit{script}, \mathcal{S}$}\tcc*[r]{Over-approximate asset transitions}
    $\Delta W \gets \text{Val}(\mathcal{B}_{final}, \Omega_{final}) - \text{Val}(\mathcal{B}_{init}, \Omega_{init})$\;
    \If{$\Delta W > 0$}{
        \Return \textbf{true}
    }
    \lElse{
        \Return \textbf{false}
    }
}

\end{algorithm}

\subsection{Exploit Synthesis with Feasibility Checking}
\label{sec:validation}

As shown in Algorithm~\ref{alg:exploit_synthesis}, this process is formalized as an iterative search for a valid exploit sequence $\mathcal{P}$ satisfying reachability and profitability constraints, instantiated into a feasible exploit script.

\textbf{Two-stage generation}.
The two stages refer to: (1) PoC plan generation, which produces a high-level, human-readable attack strategy specifying the sequence of protocol interactions, asset flows, and exploitation steps; and (2) PoC script generation, which instantiates the validated plan into a concrete, executable Solidity test script instrumented for Foundry.
Rather than relying on rigid templates, synthesis utilizes exploit primitives grounded in \sys's working memory and the LLM's generative capabilities.
It begins with an exploit plan $P = \{t_1, t_2, \dots, t_n\}$, organizing the attack into three phases: preparation (environment setup and asset acquisition), exploitation (triggering the vulnerability), and extraction (profit realization).
Each transaction $t_i$ is represented as $\langle \mathcal{C}, f, \sigma, \mathcal{K} \rangle$, where $\mathcal{C}$ is the target contract, $f$ the entrance function, $\sigma$ the parameters (concrete or symbolic), and $\mathcal{K}$ the primary target operation (state modification, external call, or fund transfer).
Once validated, the LLM integrates the plan with the execution environment $\mathcal{E}$ (ABIs, contract addresses, block heights) to synthesize a Foundry-based~\cite{Foundry} PoC script instrumented with diagnostic oracles for verifiable proof of exploit success.

\textbf{Reachability and profitability validation.}
This stage combines LLM-guided semantic reasoning with formal constraint solving to prune infeasible PoC candidates before Foundry execution.
We avoid heavyweight symbolic verification: the goal is cheap pruning that produces actionable failure signals for iterative refinement, since pure Foundry failures offer little insight into \emph{why} a candidate fails.
The two stages target the two dimensions exploits must satisfy: logical reachability and economic viability.

In the \textit{reachability validation}, for each transaction $t = \langle \mathcal{C}, f, \sigma, \mathcal{K} \rangle \in \mathcal{P}$, we evaluate whether $\mathcal{K}$ is logically reachable from $f$ at the inter-procedural constraint level.
\textsc{SemanticTraversal} performs LLM-guided traversal of $\mathcal{C}$, expanding callees, tracking how symbolic parameters $\sigma$ propagate across scopes and contract boundaries, and emitting an ordered call sequence $\pi = \langle f_0, \ldots, f_k \rangle$ with associated data-flow edges.
\textsc{CollectPredicates} then walks $\pi$ and encodes each \texttt{require}/\texttt{assert}/\texttt{branch} condition as a Z3 formula $\phi_i$, treating attacker-controlled inputs and relevant protocol state as symbolic.
A satisfying assignment for $\bigwedge \Phi$ concretizes $\sigma$ and confirms reachability; UNSAT certifies that the path cannot be executed under any input and the candidate is pruned.
The check focuses on logic within the target protocol's own contracts, where exploit-relevant predicates concentrate; behaviors of unrelated external contracts encountered along $\pi$ are assigned conservative defaults that may admit additional candidates but never spuriously reject feasible ones.
Within these bounds, an UNSAT verdict from Z3 reliably eliminates infeasible paths once $(\pi, \Phi)$ is extracted.

The \textit{profitability stage} filters candidates that are reachable but cannot yield positive net wealth.
We maintain an abstract asset state $\mathcal{S} = \langle \mathcal{B}, \Omega \rangle$, where $\mathcal{B}$ tracks token balances of the attacker and relevant contracts, and $\Omega$ captures price-relevant state from AMM reserves or external oracles.
The LLM simulates script execution at the asset layer only: recording balance and price transitions while abstracting control flow under idealized conditions such as zero slippage and successful branch completion.
A candidate proceeds to Foundry if
\begin{equation}
  \small
\Delta W = \text{Val}(\mathcal{B}_{final}, \Omega_{final}) - \text{Val}(\mathcal{B}_{init}, \Omega_{init}) > 0.
\label{eq:net_wealth_change}
\end{equation}
The filter is optimistic: idealized assumptions favor the exploitation, so the gate admits any candidate profitable under at least one plausible execution and rejects only those with no path to a positive $\Delta W$.
A walk-through of both validation stages on a flash-loan governance exploit candidate is provided in Appendix~\ref{app:walkthrough}.

\section{Implementation}
The system is implemented in Python with approximately 12K lines of code, using LangChain~\cite{Langchain} to orchestrate multi-step LLM interactions and state management. 

\textbf{Contract preprocessing} is implemented using ANTLR~\cite{ANTLR} to extract structural metadata (functions, visibility, and state variables) and generate abstract syntax trees (ASTs).
For multi-file projects, we use a lightweight call-graph–based pruning to identify core logic contracts, reducing token overhead and improving analysis stability.

\textbf{Knowledge storage and indexing}.
The hierarchical knowledge graph is implemented through a dual-storage architecture.
We use Neo4j~\cite{Neo4j} as the graph database to store the graph topology for efficient traversal and reasoning across connected nodes.
Complementary to the graph, node descriptions are indexed in Faiss~\cite{Faiss}, which serves as the RAG vector store to enable similarity-based retrieval for both initial reasoning seeds and cross-layer node discovery.

\textbf{Verification engine.}
The reachability check requires translating Solidity-level predicates into a form consumable by Z3.
We implement a normalization layer that flattens mapping accesses, resolves type casts, rewrites Solidity-specific arithmetic into Z3-compatible expressions, and applies conservative defaults for undeclared external returns.
Predicate extraction uses a structured prompt that asks the LLM to emit each branch condition in a fixed form, which the layer then parses and assembles into the SMT query.
The profitability check is realized as an LLM-driven simulator that maintains the abstract state $\mathcal{S}$ across simulated calls, with placeholder-aware handling of unresolved external values and standardized asset-accounting templates to keep $\Delta W$ comparable across heterogeneous protocols.

\section{Evaluation}

\subsection{Evaluation Setups}

\textbf{Datasets}.
\textbf{D1} is built from 72 Code4rena-audited projects in the Web3Bugs dataset collected by Sun et al.~\cite{10.1145/3597503.3639117}, comprising 2,573 smart contracts.
It includes 41 projects with 48 verified vulnerability types and 31 vulnerability-free projects.
Dataset \textbf{D2} contains 88 real-world attack cases collected from DeFiHackLabs~\cite{defiheck}, covering those used in \textsc{Verite} and \textsc{A1}.
Dataset \textbf{D3} consists of contract bounty projects from the Secure3~\cite{Secure3} platform.

\textbf{Research questions.}
Based on the above datasets, we aim to answer these research questions:
\begin{itemize}[leftmargin=4mm, itemindent=0mm]
  \item \textbf{RQ1:} How effective is \sys in identifying and exploiting real-world vulnerabilities, and how efficient is it in terms of runtime performance and token consumption?
  
  \item \textbf{RQ2:} How does \sys perform compared to state-of-the-art approaches including fuzzers as well as the LLM-based scanner and exploit generator?

  \item \textbf{RQ3:} How does hierarchical knowledge and the validation module affect \sys's performance?

  \item \textbf{RQ4:} Can \sys detect previously unknown (0-day) vulnerabilities in real-world projects?
\end{itemize}

\textbf{Baselines}.
To ensure a comprehensive evaluation, we compare \sys with four representative tools across different technical paradigms. 
For vulnerability identification, we use \textsc{GPTScan}~\cite{10.1145/3597503.3639117}, which represents the current SOTA LLM-based vulnerability scanner as our baseline. 
For vulnerability exploitation, we select \textsc{ItyFuzz}~\cite{shou2023ityfuzz}, \textsc{Verite}~\cite{VERITE}, and \textsc{A1}~\cite{gervais2025aiagentsmartcontract}. 
\textsc{ItyFuzz} and \textsc{Verite} are SOTA fuzzers capable of triggering profitable vulnerabilities, while \textsc{A1} is an LLM-based PoC generator.

\textbf{LLM selection}.
We employ five widely used LLMs as the backbone models of \sys, namely GPT-5.2, GPT-5, GPT-4o, GPT-o3, and GPT-3.5-turbo. 
These models cover a broad parameter range from tens of billions to several trillion parameters and include both general-purpose and inference-optimized variants, enabling a comprehensive evaluation of \sys across different model capacities.
For all models, we use their publicly available API endpoints with a temperature of 0.2 to ensure deterministic behavior across runs.

\begin{table}[t!]
\scriptsize
\setlength{\tabcolsep}{3.5pt}
\caption{Performance of \sys in detecting vulnerabilities within dataset D1.}
\label{tab:D1}
\begin{tabular}{@{}clcccccccc@{}}
\toprule
\multirow{2}{*}{\textbf{Dataset}} &
  \multicolumn{1}{c}{\multirow{2}{*}{\textbf{Model}}} &
  \multicolumn{8}{c}{\textbf{Result}} \\ \cmidrule(l){3-10} 
 &
  \multicolumn{1}{c}{} &
  \textbf{TP} &
  \textbf{FN} &
  \textbf{FP} &
  \textbf{TN} &
  \textbf{Precision} &
  \textbf{Accuracy} &
  \textbf{Recall} &
  \textbf{F1-score} \\ \midrule
\multirow{4}{*}{D1} & GPT-5   & 47 & 1 & 10 & 29 & 0.82 & 0.87 & 0.98 & 0.90 \\
                    & GPT-4o  & 44 & 4 & 13 & 25 & 0.77 & 0.80 & 0.92 & 0.84 \\
                    & GPT-o3  & 47 & 1 & 12 & 25 & 0.80 & 0.85 & 0.98 & 0.88 \\
                    & GPT-3.5 & 45 & 3 & 15 & 27 & 0.75 & 0.80 & 0.94 & 0.83 \\ \bottomrule
\end{tabular}
\end{table}

\begin{table*}[t!]
  \setlength{\tabcolsep}{5.5pt}
\scriptsize
\centering
\caption{Exploitation performance and revenue of \sys on Dataset D2. ESR denotes the Exploit Success Rate. Asterisks (*) denote projects where PoC failed to achieve exploitation.}
\label{tab:D2}
\begin{tabular}{@{}c|lrcr|c|lrcr|c|lrcr@{}}
\toprule
\multicolumn{1}{c|}{\textbf{ID}} &
  \multicolumn{1}{c}{\textbf{Name}} &
  \multicolumn{1}{c}{\textbf{\begin{tabular}[c]{@{}c@{}}Max\\ Revenue\end{tabular}}} &
  \textbf{\begin{tabular}[c]{@{}c@{}}Avg\\ Iter\end{tabular}} &
  \multicolumn{1}{c|}{\textbf{\begin{tabular}[c]{@{}c@{}}Avg\\ Token\end{tabular}}} &
  \multicolumn{1}{c|}{\textbf{ID}} &
  \multicolumn{1}{c}{\textbf{Name}} &
  \multicolumn{1}{c}{\textbf{\begin{tabular}[c]{@{}c@{}}Max\\ Revenue\end{tabular}}} &
  \textbf{\begin{tabular}[c]{@{}c@{}}Avg\\ Iter\end{tabular}} &
  \multicolumn{1}{c|}{\textbf{\begin{tabular}[c]{@{}c@{}}Avg\\ Token\end{tabular}}} &
  \multicolumn{1}{c|}{\textbf{ID}} &
  \multicolumn{1}{c}{\textbf{Name}} &
  \multicolumn{1}{c}{\textbf{\begin{tabular}[c]{@{}c@{}}Max\\ Revenue\end{tabular}}} &
  \textbf{\begin{tabular}[c]{@{}c@{}}Avg\\ Iter\end{tabular}} &
  \multicolumn{1}{c}{\textbf{\begin{tabular}[c]{@{}c@{}}Avg\\ Token\end{tabular}}} \\ \midrule
1 &
  aes &
  \$61.6K &
  1.3 &
  148.9K &
  31 &
  zeed &
  \$124.5K &
  4.0 &
  106.0K &
  61 &
  opyn &
  \$9.9K &
  2.3 &
  293.6K \\
2 &
  apemaga &
  \$57.4K &
  2.3 &
  61.0K &
  32 &
  bno &
  \$0.0K &
  3.3 &
  132.0K &
  62 &
  allbridge &
  \$5.5K &
  1.8 &
  380.6K \\
3 &
  axioma &
  \$18.9K &
  2.0 &
  69.5K &
  33 &
  curve01 &
  \$2,504.5K &
  1.8 &
  136.4K &
  63 &
  annex &
  \$6.6K &
  3.5 &
  182.8K \\
4 &
  bamboo &
  \$205.2K &
  1.5 &
  148.5K &
  34 &
  cover &
  \$1,274.4K &
  1.0 &
  81.3K &
  64 &
  apedao &
  \$7.5K &
  1.3 &
  132.3K \\
5 &
  bego &
  \$10.9K &
  1.5 &
  156.2K &
  35 &
  newfi &
  \$30.5K &
  3.0 &
  143.1K &
  65 &
  bigfi &
  \$30.3K &
  2.5 &
  164.1K \\
6 &
  bevo &
  \$130.7K &
  1.8 &
  158.7K &
  36 &
  utopia &
  \$446.6K &
  2.3 &
  105.7K &
  66 &
  cs &
  \$684.2K &
  2.0 &
  168.3K \\
7 &
  bunn &
  \$47.2K &
  3.5 &
  130.7K &
  34 &
  wgpt &
  \$76.9K &
  4.5 &
  95.0K &
  67 &
  depusdt &
  \$106.1K &
  1.8 &
  72.1K \\
8 &
  cellframe &
  \$222.8K &
  3.8 &
  49.8K &
  38 &
  myai &
  \$9.8K &
  2.5 &
  65.4K &
  68 &
  discover* &
  \$0.0K &
  5.0 &
  122.1K \\
9 &
  dfs &
  \$1.5K &
  3.5 &
  115.1K &
  39 &
  ddcoin &
  \$126.4K &
  3.8 &
  80.3K &
  69 &
  dpc &
  \$10.8K &
  3.0 &
  223.0K \\
10 &
  fapen &
  \$10.9K &
  1.5 &
  45.9K &
  40 &
  cfc &
  \$19.1K &
  3.8 &
  116.8K &
  70 &
  gds &
  \$207.2K &
  4.0 &
  154.9K \\
11 &
  fil314 &
  \$12.8K &
  4.0 &
  74.2K &
  41 &
  babydoge &
  \$401.1K &
  4.3 &
  170.7K &
  71 &
  gym\_1 &
  \$1,246.6K &
  2.0 &
  61.9K \\
12 &
  game &
  \$99.0K &
  2.0 &
  58.1K &
  42 &
  bzx &
  \$9,486.7K &
  2.8 &
  94.5K &
  72 &
  hackdao &
  \$148.5K &
  2.3 &
  112.0K \\
13 &
  gss &
  \$24.9K &
  3.8 &
  248.9K &
  43 &
  mamo &
  \$5.3K &
  3.5 &
  99.7K &
  73 &
  inuko &
  \$5,019.5K &
  2.3 &
  135.6K \\
14 &
  health &
  \$15.1K &
  2.3 &
  131.5K &
  44 &
  hypr &
  \$1.4K &
  3.5 &
  65.0K &
  74 &
  lusd &
  \$9.5K &
  1.8 &
  69.1K \\
15 &
  hpay &
  \$103.9K &
  3.3 &
  52.0K &
  45 &
  compound &
  \$45.1K &
  4.5 &
  330.7K &
  75 &
  lw &
  \$83.5K &
  4.0 &
  137.0K \\
16 &
  mbc &
  \$5.9K &
  4.0 &
  144.1K &
  46 &
  ffist &
  \$207.2K &
  1.3 &
  107.5K &
  76 &
  neverfall &
  \$74.3K &
  3.0 &
  177.1K \\
17 &
  melo &
  \$250.4K &
  1.0 &
  58.5K &
  47 &
  laeeb &
  \$43.2K &
  1.5 &
  176.2K &
  77 &
  omniestate &
  \$0.1K &
  1.0 &
  131.3K \\
18 &
  olife &
  \$29.3K &
  3.5 &
  116.4K &
  48 &
  juice &
  \$98.3K &
  2.3 &
  59.6K &
  78 &
  res &
  \$184.4K &
  2.5 &
  152.6K \\
19 &
  pledge &
  \$15.0K &
  1.5 &
  198.0K &
  49 &
  smartmesh &
  \$0.5K &
  2.5 &
  66.8K &
  79 &
  roi &
  \$152.5K &
  2.0 &
  198.5K \\
20 &
  pltd &
  \$244.5K &
  1.0 &
  102.5K &
  50 &
  spankchain &
  \$506.0K &
  2.8 &
  146.1K &
  80 &
  safemoon* &
  \$0.0K &
  5.0 &
  112.1K \\
21 &
  rfb &
  \$5.6K &
  2.0 &
  102.1K &
  51 &
  yearn\_ydai &
  \$185.1K &
  4.8 &
  58.8K &
  81 &
  sdao &
  \$13.7K &
  1.3 &
  106.9K \\
22 &
  seama &
  \$7.8K &
  2.0 &
  149.5K &
  52 &
  spartan &
  \$1,529.4K &
  1.8 &
  57.9K &
  82 &
  selltoken &
  \$10.9K &
  2.8 &
  74.1K \\
23 &
  shadowfi &
  \$978.9K &
  1.0 &
  150.6K &
  53 &
  bearn &
  \$123.1K &
  4.0 &
  78.7K &
  83 &
  sheep &
  \$20.0K &
  2.3 &
  162.3K \\
24 &
  sut &
  \$29.6K &
  1.5 &
  56.4K &
  54 &
  hunny &
  \$5.2K &
  4.0 &
  546.8K &
  84 &
  sheepfarm &
  \$0.1K &
  3.0 &
  101.4K \\
25 &
  swapos &
  \$9.5K &
  1.8 &
  68.3K &
  55 &
  popsicle &
  \$2.2K &
  4.8 &
  404.0K &
  85 &
  starlink &
  \$34.8K &
  1.3 &
  169.0K \\
26 &
  uerii &
  \$5.9K &
  1.0 &
  80.1K &
  56 &
  nimbus &
  \$4.6K &
  3.5 &
  92.1K &
  86 &
  tinu &
  \$70.0K &
  2.3 &
  243.8K \\
27 &
  upswing &
  \$1.0K &
  3.8 &
  59.2K &
  57 &
  ploutoz* &
  \$0.0K &
  5.0 &
  57.5K &
  87 &
  ufdao &
  \$227.1K &
  3.3 &
  67.1K \\
28 &
  uranium &
  \$8,772.6K &
  3.0 &
  69.9K &
  58 &
  yeth &
  \$1,077.0K &
  4.5 &
  183.9K &
  88 &
  valuedefi &
  \$359.6K &
  1.3 &
  100.6K \\
29 &
  uwerx &
  \$63,797.5K &
  1.5 &
  102.1K &
  59 &
  drlvaultv3 &
  \$13,980.8K &
  3.5 &
  127.2K &
   &
   &
   &
   &
   \\
30 &
  wifcoin &
  \$10.8K &
  2.3 &
  141.0K &
  60 &
  bancor &
  \$0.2K &
  2.0 &
  187.2K &
   &
   &
   &
   &
   \\ \midrule
\textbf{Total} &
  \multicolumn{2}{l}{\textbf{Exploted: 85/88}} &
  \multicolumn{2}{l}{\textbf{ESR:96.6\%}} &
  \multicolumn{10}{l}{\textbf{Revenue: \$116,225.3K}} \\ \bottomrule
\end{tabular}
\end{table*}

\subsection{RQ1: Effectiveness in Real-world Scenarios}
We evaluate \sys on datasets D1 and D2 under different LLMs as the base model.
D1 consists of audited projects and is used to evaluate \sys's effectiveness in vulnerability identification.
D2 contains historical Defi attack incidents with confirmed profits as ground truth and is used to assess whether the PoCs generated by \sys can reproduce the exploits and yield profits.

\begin{table*}[t!]
\centering
\scriptsize
\setlength{\tabcolsep}{10pt}
\caption{Exploitation performance of \sys on Dataset D2 across different underlying LLMs.}
\label{tab:ExploitationLLM}
\begin{threeparttable}
\begin{tabular}{@{}clcccccccr@{}}
\toprule
\multirow{2.5}{*}{\textbf{Dataset}} &
\multicolumn{1}{c}{\multirow{2.5}{*}{\textbf{Model}}} &
\multicolumn{8}{c}{\textbf{Result}} \\ 
\cmidrule(l){3-10}
 &
\multicolumn{1}{c}{} &
\textbf{Det.} &
\textbf{Rec.} &
\textbf{Exp.} &
\textbf{ESR} &
\textbf{\begin{tabular}[c]{@{}c@{}}Avg.\\ Iteration\end{tabular}} &
\textbf{\begin{tabular}[c]{@{}c@{}}Avg.\\ Time\end{tabular}} &
\textbf{\begin{tabular}[c]{@{}c@{}}Avg.\\ Token\end{tabular}} &
\multicolumn{1}{c}{\textbf{\begin{tabular}[c]{@{}c@{}}Total\\ Revenue\end{tabular}}} \\
\midrule
\multirow{4}{*}{D2} 
& GPT-5.2 & 82 & 93.2\% & 76 & 86.4\% & 2.84 & 296.50 & 151,278 & \$110,960,607.9 \\
& GPT-5   & 82 & 93.2\% & 69 & 78.4\% & 2.32 & 119.48 & 114,889 & \$107,626,754.2 \\
& GPT-o3  & 85 & 96.6\% & 79 & 89.8\% & 2.49 & 178.94 & 142,827 & \$86,384,234.5  \\
& GPT-4o  & 77 & 87.5\% & 70 & 79.5\% & 2.86 & 100.69 & 119,502 & \$94,936,898.2  \\
\bottomrule
\end{tabular}%
\begin{tablenotes}
\footnotesize
\item Det., Exp., and Rec. denote the number of detected vulnerabilities, exploited vulnerabilities, and recall rate, respectively.
\end{tablenotes}
\end{threeparttable}
\end{table*}

\textbf{Vulnerability detection performance (D1)}.
Table~\ref{tab:D1} summarizes the vulnerability detection performance of \sys on D1.
TP and FN denote the numbers of detected and missed vulnerabilities among the 48 kinds of vulnerabilities in D1.
FP denotes the false alarms during \sys's analysis, and TN denotes benign projects correctly identified as non-vulnerable.
Overall, \sys demonstrates strong detection capability across all base models, achieving high recall (up to 0.98) and F1-scores between 0.83 and 0.90.
Notably, with GPT-5 as the base model, \sys detects 47 out of 48 vulnerability types while maintaining a moderate false positive rate.

This high recall stems from the HKG's contract semantics layer, which encodes protocol-level behavioral patterns that pure static analysis typically misses. 
Rather than relying on predefined syntactic rules, \sys retrieves semantically relevant protocol and dependency nodes and performs multi-hop reasoning to connect localized code anomalies to broader vulnerability contexts, enabling detection of complex logic flaws that manifest only through cross-contract or cross-function interactions. 
The moderate false positive rate reflects two failure modes. 
First, \sys occasionally misinterprets intentionally permissive access-control designs as vulnerabilities, as the underlying developer intent is not always recoverable from code alone. 
Second, some potential vulnerabilities are not practically exploitable, such as reentrancy where existing guards or rollback mechanisms prevent actual exploit.

\textbf{Real-world exploitation performance (D2)}.
Table~\ref{tab:D2} presents the exploitation results and revenue achieved by \sys on dataset D2, which consists of real-world attack incidents spanning from 2022 to December 2025.
Overall, the PoCs generated by \sys successfully reproduced 85 out of 88 historical exploits, yielding an exploit success rate (ESR) of 96.6\% and a total reproduced revenue of \$116,225.3K.
The largest reproduced profit is observed on \textit{uwerx}, reaching \$63,797.5K, which exceeds the profit obtained by the real-world attacker in the original incident.
The iteration refers to the repetitive generation process triggered when the initial PoC fails our two-step verification, and a task is marked as a failure if it exceeds a maximum of 5 iterations.
The strong ESR reflects several complementary design choices.
HKG provides semantic context, links vulnerable logic to root causes, and offers high-level exploit guidance, enabling \sys to compose structured PoCs grounded in real-world precedents rather than generating exploit code from scratch.
The two-stage validation further filters structurally infeasible and economically non-viable candidates before Foundry execution, focusing refinement on plausible PoCs and providing informative feedback for iterative repair.

\begin{figure*}[t!]
  \centering
  \includegraphics[width=5.6in]{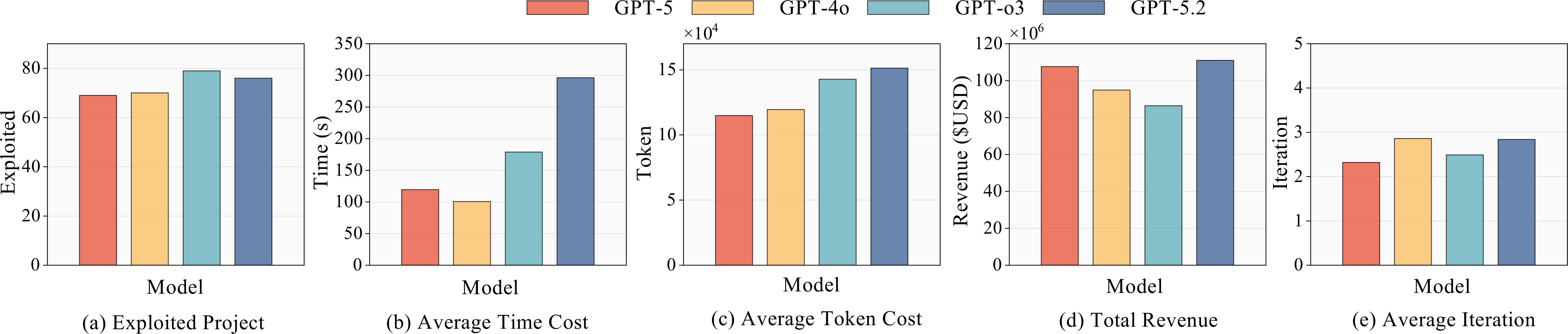}
  \caption{Performance metrics of \sys across various LLM backends.}
  \label{fig:Performance}
\end{figure*}

\begin{figure*}[t!]
  \centering
  \includegraphics[width=5.6in]{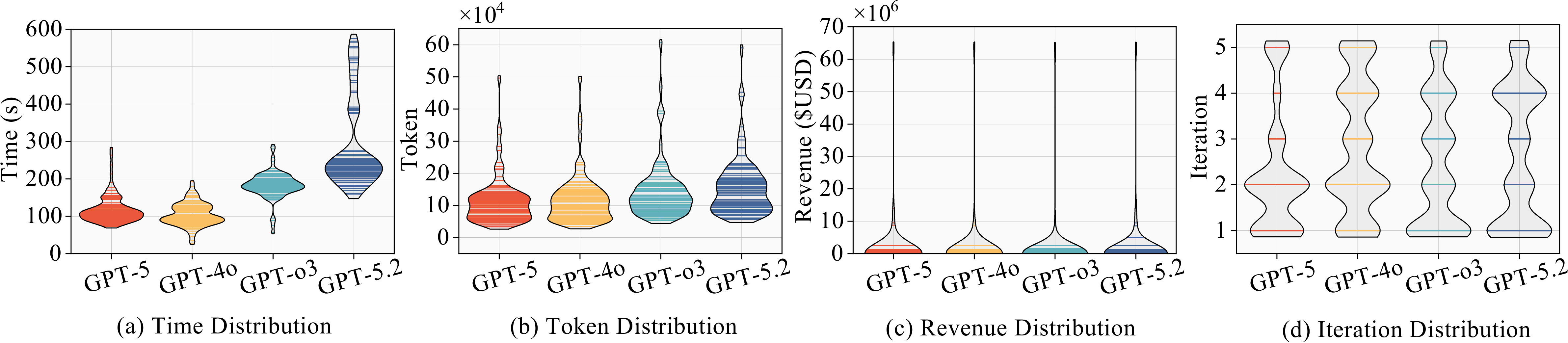}
  \caption{Distribution analysis of performance indicators for each model.}
  \label{fig:Distribution}
\end{figure*}

We also analyze the three failed exploit cases. 
In the project \textit{discover}, \sys triggers the unfair exchange rate but fails to construct a profit-realizing multi-contract exploit due to ambiguity between price-manipulation logic and arithmetic errors. 
In project \textit{ploutoz}, the vulnerability lies in low-level \texttt{mint}/\texttt{burn} arithmetic corner cases unrelated to function semantics, causing \sys to miss the relevant execution path. 
In project \textit{safemoon}, publicly exposed \texttt{mint} and \texttt{burn} interfaces in a proxy contract are vulnerable, but \sys focuses on proxy risks, leading to an incorrect contract-type abstraction and missed trigger.

\textbf{Cross-LLM effectiveness and efficiency analysis}.
We break down \sys's performance on D2 across LLMs and summarize it in Table~\ref{tab:ExploitationLLM} and visualized in Figures~\ref{fig:Performance} and \ref{fig:Distribution}.

i) For \textit{exploitation effectiveness}, the results show that \sys is effective and robust across all evaluated LLMs. 
All models achieve an ESR above 78\% and identify more than \$86M in total revenue, indicating that the framework consistently bridges the gap between raw LLM reasoning and practical exploit generation.
Among the models, GPT-o3 achieves the highest ESR (89.8\%), benefiting from its reasoning-oriented architecture which exhibits stronger capability in HKG knowledge retrieval.
GPT-5.2 performs best in high-impact cases, reaching the highest total Revenue of \$110M, which can be attributed to its stronger state modeling and strategy optimization ability.

ii) For \textit{efficiency and resource overhead}, GPT-4o is the most time-efficient among all models, requiring only 100.69s per task on average. 
GPT-5 is the most resource-economical, utilizing the lowest iterations (2.32) and tokens (114,889). 
In contrast, GPT-5.2 incurs the heaviest overhead, with the longest time cost (296.50s) and highest token consumption (151,278).
This trade-off indicates that stronger models improve the ability to exploit more vulnerabilities with more profits but require more extensive reasoning, leading to higher runtime and token consumption.

iii) For \textit{distributional characteristics}, in Figure~\ref{fig:Distribution} (a) and (b), models exhibit a long-tail distribution, reflecting its extensive reasoning process when handling highly complex projects.
In Figure~\ref{fig:Distribution} (c), all models follow a power-law distribution where a small number of incidents account for the majority of total revenue, reflecting the reality of DeFi security incidents.
GPT-5 and GPT-o3 show tighter distributions in Figure~\ref{fig:Distribution} (d), with most PoCs generated within 1–2 iterations, demonstrating higher determinism compared to the wider distributions of GPT-4o and GPT-5.2.

\begin{center}
  \small
  \fcolorbox{black}{gray!10}{\parbox{.97\linewidth}{\leftskip=0.6em \rightskip=0.6em \textbf{Answer to RQ1: } 
  \sys is highly effective, achieving a 0.98 recall and 0.90 F1-score in detection, and a 96.6\% exploit success rate with \$116.2M revenue. 
  It also maintains consistent performance across various LLM backends with moderate overhead in terms of runtime and token consumption.
  }}
\end{center}

\begin{table}[]
\setlength{\tabcolsep}{3pt}
\centering
\scriptsize
\caption{Comparison of vulnerability detection performance between \sys and \textsc{GPTScan} on D1.}
\label{tab:Comparison1}
\begin{tabular}{@{}clccccc@{}}
\toprule
\multirow{2}{*}{\textbf{Dataset}} & \multicolumn{1}{c}{\multirow{2}{*}{\textbf{Tool}}} & \multirow{2}{*}{\textbf{Model}} & \multicolumn{4}{c}{\textbf{Result}} \\ \cmidrule(l){4-7} 
                    & \multicolumn{1}{c}{}   &         & \textbf{Precision} & \textbf{Accuracy} & \textbf{Recall} & \textbf{F1-score} \\ \midrule
\multirow{2}{*}{D1} & \textsc{GPTScan}                & GPT-3.5 & 57.1\%               & 62.4\%              & 83.3\%            & 67.8\%              \\
                    & \textbf{\sys (Ours)} & GPT-3.5 & \textbf{75.0\%}      & \textbf{80.0\%}     & \textbf{93.8\%}   & \textbf{83.3\%}     \\ \bottomrule
\end{tabular}

\end{table}

\subsection{RQ2: Comparison with SOTA Approaches}

\textbf{Comparison with vulnerabiliy scanner}.
Table~\ref{tab:Comparison1} reports the performance of \sys and \textsc{GPTScan} on the D1 dataset. 
\textsc{GPTScan} achieves a high recall of 0.83 but suffers from relatively low precision (0.57), indicating a considerable number of false positives. 
Specifically, it detects 40 out of 48 vulnerability types while missing 8 cases and producing 30 false positives. 
In contrast, \sys consistently outperforms \textsc{GPTScan} across all evaluation metrics, achieving higher precision (0.75), accuracy (0.80), recall (0.94), and F1-score (0.83). 
These results demonstrate that \sys leverages its memory mechanism and knowledge-based reasoning chains to substantially mitigate LLM hallucinations, resulting in fewer false positives and more dependable detection.

\textbf{Comparison with exploitation generator}.
Our dataset D2 encapsulates the benchmarks used by compared tools.
Specifically, we derived D2 (s1) by incorporating all incidents from \textsc{ItyFuzz} and \textsc{Verite}, retaining 52 cases after excluding those where source code is no longer available. 
Similarly, D2 (s2) includes the incidents from the \textsc{A1} benchmark, with 31 cases remaining.
As shown in Table~\ref{tab:compareD2}, \sys achieves an ESR of 96.2\% and 93.5\% on D2 (s1) and D2 (s2), nearly 2$\times$ of \textsc{Verite}’s 48.1\% and \textsc{A1}'s 48.4\%, and 5$\times$ of \textsc{ItyFuzz}.
\sys also demonstrates a massive lead in profitability.
Compared to its LLM-based counterpart \textsc{A1}, it achieves 8.5$\times$ the total revenue and 4.4$\times$ the average revenue.
\textsc{Verite} achieves a slightly higher average revenue by employing a profit maximizer. 
In contrast, \sys generates PoCs through knowledge-driven logic reasoning without specific optimization for profit maximization.

\begin{table}[t!]
\setlength{\tabcolsep}{2pt}
\centering
\scriptsize
\caption{Exploitation performance comparison between \sys and SOTA fuzzers and PoC generator on D2.}
\label{tab:compareD2}
\begin{tabular}{@{}clcccrr@{}}
\toprule
\multirow{2.5}{*}{\textbf{Dataset}} &
  \multicolumn{1}{c}{\multirow{2.5}{*}{\textbf{Tool}}} &
  \multirow{2.5}{*}{\textbf{Model}} &
  \multicolumn{4}{c}{\textbf{Result}} \\ \cmidrule(l){4-7} 
 &
  \multicolumn{1}{c}{} &
   &
  \textbf{Exploited} &
  \textbf{ESR} &
  \textbf{Total Revenue} &
  \textbf{Average Revenue} \\ \midrule
\multirow{3}{*}{D2 (s1)} & \textsc{ItyFuzz}       & \multirow{3}{*}{N\textbackslash{}A} & 10 & 19.2\% & \$106,473.8    & \$10,647.4    \\
                        & \textsc{Verite}         &                                     & 25 & 48.1\% & \$18,225,320.8 & \$729,012.8   \\
                        & \textbf{\sys} &                                     & \textbf{50} & \textbf{96.2\%} & \textbf{\$35,308,169.5} & \textbf{\$706,163.4}   \\ \midrule
\multirow{2}{*}{D2 (s2)} & \textsc{A1}            & GPT-o3                              & 15 & 48.4\% & \$8,839,461.0  & \$589,297.4   \\
                        & \textbf{\sys} & GPT-o3                              & \textbf{29} & \textbf{93.5\%} & \textbf{\$75,018,779.7} & \textbf{\$2,586,854.5} \\ \bottomrule
\end{tabular}

\end{table}

\begin{center}
  \small
  \fcolorbox{black}{gray!10}{\parbox{.97\linewidth}{\leftskip=0.6em \rightskip=0.6em \textbf{Answer to RQ2: }
  \sys significantly outperforms SOTA baselines, surpassing \textsc{GPTScan} with an 0.83 F1-score, 2$\times$ the exploit success rate of \textsc{Verite} and \textsc{A1}, and 8.5$\times$ the total revenue of \textsc{A1}, proving its superior effectiveness in both vulnerability detection and high-profit exploit generation.
  }}
\end{center}

\begin{table*}[t!]
  \caption{Ablation study results. CS, FM, and EP denote Contract Semantics, Failure Mode, and Exploit Primitive. Num denotes the successful exploits, and Degr denotes the degradation over baseline.}
  \label{tab:Ablation}
  \setlength{\tabcolsep}{6pt}
  \centering
\scriptsize
\begin{tabular}{@{}lcccccrcrrrrr@{}}
\toprule
\multicolumn{1}{c}{\multirow{3}{*}{\textbf{Tool}}} &
  \multirow{3}{*}{\textbf{Model}} &
  \multicolumn{3}{c}{\multirow{2}{*}{\textbf{Knowledge}}} &
  \multicolumn{8}{c}{\textbf{Result}} \\
\multicolumn{1}{c}{} &
   &
  \multicolumn{3}{c}{} &
  \multicolumn{2}{c}{\textbf{Detected}} &
  \multicolumn{2}{c}{\textbf{Exploited}} &
  \multicolumn{2}{c}{\textbf{Total Revenue}} &
  \multicolumn{2}{c}{\textbf{Average Revenue}} \\ \cmidrule(l){3-5}\cmidrule(l){6-7}\cmidrule(l){8-9}\cmidrule(l){10-11}\cmidrule(l){12-13}
\multicolumn{1}{c}{} &
   &
  \textbf{CS} &
  \textbf{FM} &
  \textbf{EP} &
  \textbf{Num} &
  \multicolumn{1}{c}{\textbf{Degr}} &
  \textbf{Num} &
  \multicolumn{1}{c}{\textbf{Degr}} &
  \multicolumn{1}{c}{\textbf{Num}} &
  \multicolumn{1}{c}{\textbf{Degr}} &
  \multicolumn{1}{c}{\textbf{Num}} &
  \multicolumn{1}{c}{\textbf{Degr}} \\ \midrule
\sys-SF &
  GPT-5 &
  \color{red}{$\times$} &
  \color{red}{$\times$} &
  \color{mygreen}{$\checkmark$} &
  17 &
  \color{mygreen}{-79\%} &
  11 &
  \color{mygreen}{-84\%} &
  \$2,465,360 &
  \color{mygreen}{-98\%} &
  \$224,123 &
  \color{mygreen}{-86\%} \\
\sys-EP &
  GPT-5 &
  \color{mygreen}{$\checkmark$} &
  \color{mygreen}{$\checkmark$} &
  \color{red}{$\times$} &
  80 &
  \color{mygreen}{-2\%} &
  1 &
  \color{mygreen}{-99\%} &
  \$931,177 &
  \color{mygreen}{-99\%} &
  \$931,177 &
  \color{mygreen}{-40\%} \\
\sys-NO &
  GPT-5 &
  \color{red}{$\times$} &
  \color{red}{$\times$} &
  \color{red}{$\times$} &
  16 &
  \color{mygreen}{-80\%} &
  0 &
  \color{mygreen}{-100\%} &
  0 &
  \color{mygreen}{-100\%} &
  0 &
  \color{mygreen}{-100\%} \\
\textbf{\sys} &
  GPT-5 &
  \color{mygreen}{$\checkmark$} &
  \color{mygreen}{$\checkmark$} &
  \color{mygreen}{$\checkmark$} &
  \textbf{82} &
  \multicolumn{1}{r}{N/A} &
  \textbf{69} &
  \multicolumn{1}{r}{N/A} &
  \textbf{\$107,626,754} &
  \multicolumn{1}{r}{N/A} &
  \textbf{\$1,559,808} &
  \multicolumn{1}{r}{N/A} \\ \bottomrule
\end{tabular}%

\end{table*}

\subsection{RQ3: Ablation Study}
\label{sec_ab}

We conduct ablation studies to assess the contribution of \sys's knowledge layers and validation module to vulnerability detection, exploit generation, and refinement efficiency.

\textbf{HKG ablation}.
We construct three variants by selectively removing HKG layers: \sys-SF removes the contract semantics and failure mode layers, \sys-EP removes the exploit primitive layer, and \sys-NO removes all layers.
As shown in Table~\ref{tab:Ablation}, all layers are necessary for end-to-end exploit synthesis.
Removing the semantic and failure mode layers (\sys-SF) reduces detection by 79\% and total revenue by 98\%.
This suggests that exploit primitives alone are insufficient: without semantic context and failure-mode guidance, the model struggles to identify protocol-specific vulnerable logic and thus misses most exploitable cases.
In contrast, removing the exploit primitive layer (\sys-EP) retains 80 detected vulnerabilities but reduces successful exploits by 99\%.
This indicates that semantic and failure-mode knowledge mainly support vulnerability discovery, whereas exploit primitives are critical for transforming detected flaws into executable PoCs.
Removing all layers (\sys-NO) leads to zero successful exploits, further confirming that the three layers jointly bridge code understanding, vulnerability reasoning, and exploit construction.

\textbf{Validation ablation}.
Table~\ref{tab:Validation} evaluates the reachability-profitability validation module from two aspects.
Panel (a) shows that two-stage validation reduces average Foundry iterations from 13.8 to 2.3 and runtime from 309s to 117s, despite introducing 4.3 lightweight internal iterations.
This suggests that the dominant cost in refinement comes from coarse execution-level feedback.
In direct-to-Foundry refinement, a failed PoC only reveals that the complete script does not work, but does not isolate whether the failure comes from an infeasible path, wrong parameters, incorrect transaction ordering, or non-profitable asset movement.
The model therefore tends to revise the script by trial and error, often remaining in the same invalid search region.
Two-stage validation changes this process by moving part of the refinement to the plan level.
Reachability and profitability checks expose failures before script execution and provide more localized correction signals.
Thus, the additional internal iterations are cheaper and more informative, replacing many expensive trial-error attempts.

Panel (b) shows that validation removes all 8 false positives while preserving all 22 true positives.
This is because LLM-generated false positives are often not arbitrary mistakes, but partial exploits that are semantically plausible yet violate necessary exploit conditions.
For example, a candidate may match a known vulnerability pattern but fail to reach the target operation, or it may reach relevant logic without yielding positive net wealth.
By enforcing path feasibility and economic gain before Foundry execution, the validation module filters these structurally invalid candidates early.
Therefore, it improves precision by reducing the number of plausible-but-invalid PoCs entering the execution loop, rather than by simply duplicating Foundry's role.

\begin{table}[t!]
\scriptsize
\caption{Ablation of the reachability-profitability validation module on D2 with GPT-5.}
\label{tab:Validation}

\centering
\textbf{(a) Efficiency vs. direct Foundry}\\
\vspace{2pt}
\setlength{\tabcolsep}{15pt}
\begin{tabular}{@{}lccc@{}}
\toprule
Setting & Int.Iter & Fdry.Iter & Time \\ \midrule
Direct-to-Foundry & 0 & 13.8 & 309s \\
Two-stage & 4.3 & 2.3 & 117s \\ \bottomrule
\end{tabular}

\vspace{8pt}

\textbf{(b) Filtering quality}\\
\vspace{2pt}
\setlength{\tabcolsep}{10pt}
\begin{tabular}{@{}lccccc@{}}
\toprule
Setting & Passed & TP & FP & FN & Prec. \\ \midrule
No validation & 30 & 22 & 8 & -- & 73\% \\
Two-stage & 22 & 22 & 0 & 0 & 100\% \\ \bottomrule
\end{tabular}

\end{table}

\begin{center}
  \small
  \fcolorbox{black}{gray!10}{\parbox{.97\linewidth}{\leftskip=0.6em \rightskip=0.6em \textbf{Answer to RQ3: }
  Each HKG layer is indispensable: semantic layers drive vulnerability detection while exploit primitives enable successful PoC generation. 
  The two-stage validation module further improves synthesis by moving refinement from expensive Foundry-level trial-and-error to more diagnostic reachability-profitability checks.
  }}
\end{center}

\subsection{RQ4: Discovery of 0-day Vulnerabilities}

\begin{table*}[t]
\setlength{\tabcolsep}{8pt}
\centering
\scriptsize
\caption{0-day vulnerabilities found by \sys}
\label{tab:0day}
\begin{tabular}{@{}cllcccc@{}}
\toprule
\textbf{ID} &
  \multicolumn{1}{c}{\textbf{Location}} &
  \multicolumn{1}{c}{\textbf{Description}} &
  \textbf{Risk} &
  \textbf{Status} &
  \textbf{Project} &
  \textbf{TVL} \\ \midrule
1 &
  \_calculateReward() &
  Precision loss from integer truncation causing zero rewards. &
  M &
  Fixed &
  \multirow{2}{*}{cudis\_bsc} &
  \multirow{2}{*}{\$839K} \\
2 &
  emergencyWithdraw() &
  Missing blacklist check allows sanction bypass. &
  M &
  Fixed &
   &
   \\ \midrule
3 &
  removeSourceToken() &
  Reflexive token check causes DoS. &
  M &
  Fixed &
  \multirow{3}{*}{\begin{tabular}[c]{@{}c@{}}zklink \\ MergeToken\end{tabular}} &
  \multirow{3}{*}{\$37.6M} \\
4 &
  deposit() &
  Ignored transfer fees cause over-minting. &  
  M &
  Fixed &
   &
   \\
5 &
  withdraw() &
  Transfer-tax tokens break accounting, causing inflated balance. &

  M &
  Fixed &
   &
   \\ \midrule
6 &
  openEnvelope() &
  Missing signature verification enables replay and fund theft. &
  H &
  Fixed &
  \multirow{7}{*}{\begin{tabular}[c]{@{}c@{}}Aki \\ Protocol\end{tabular}}&
  \multirow{7}{*}{\$128K} \\
7 &
  openEnvlope() &
  Reentrancy with ERC777 tokens allows repeated withdrawals. &
  H &
  Fixed &
   &
   \\
8 &
  addEnvelope() &
  Missing ID existence check allows overwrite and fund theft. &
  M &
  Fixed &
   &
   \\

9 &
  moneyThisOpen() &
  Bad randomness enables MEV manipulation. &
  M &
  Fixed &
   &
   \\
10 &
  openEnvlope() &
  Reentrancy allows repeated unauthorized withdrawals. &
  M &
  Fixed &
   &
   \\
11 &
  addEnvelope() &
  Front-running ID registration causes user DoS. &
  L &
  ACK &
   &
   \\ \midrule
12 &
  unwhitelistTarget() &
  Incorrect mapping update prevents proper whitelist removal. &
  M &
  ACK &
  \multirow{2}{*}{Klydo} &
  \multirow{2}{*}{Unk.} \\
13 &
  getTarget() &
  Calldata decoding flaw bypasses whitelist checks. &
  L &
  ACK &
   &
   \\ \midrule
14 &
  init\_insurance() &
  Missing init guard allows state overwrite and manipulation. &
  M &
  ACK &
  \multirow{3}{*}{\begin{tabular}[c]{@{}c@{}}Cooking \\ City\end{tabular}} &

  \multirow{3}{*}{\$32M} \\
15 &
  transfer() &
  Fee precision loss enables fee evasion via dust attacks. &
  M &
  ACK &
   &
   \\
16 &
  init\_config() &
  Unprotected initialization allows admin takeover. &
  L &
  ACK &
   &
   \\ \bottomrule
\end{tabular}%
\end{table*}

We apply \sys to the open bug bounty program of Secure3~\cite{Secure3}, aiming to evaluate whether \sys, equipped with specialized domain knowledge, can identify previously undiscovered 0-day vulnerabilities and validate them through proofs of concept like experienced human experts.

\textbf{Overall results}.
\sys identified a total of 21 0-day vulnerabilities across five projects. 
To date, 16 of these vulnerabilities were either confirmed or fixed by the developers, resulting in a total bounty reward of \$2,900.
As shown in Table~\ref{tab:0day}, these 16 vulnerabilities include 2 high-severity, 11 medium-severity, and 3 low-severity issues.
The detected vulnerabilities span various categories, such as accounting errors, unprotected initialization, and reentrancy. 
Notably, all high-severity and the majority of medium-severity vulnerabilities have been remediated by the developers, securing approximately \$70.6M in on-chain assets.

\renewcommand{\arraystretch}{0.85}
\begin{table*}[t]
\centering
\scriptsize
\caption{Key nodes retrieved from HKG during \sys's analysis of the fee-on-transfer token incompatibility vulnerability.}
\label{tab:case}

\begin{tabular}{p{0.23\textwidth} p{0.23\textwidth} p{0.23\textwidth}}
\toprule
\midrule

\textit{\textbf{Contract Semantics}} \newline
\textbar{} \newline
\textbar{}\hspace{0.5em}+-- \textit{\textbf{Protocol}} \newline
\textbar{}\hspace{1.5em}+-- \textit{TokenWrapping} \newline
\textbar{}\hspace{2.5em}+-- \textit{Portal} \newline
\textbar{}\hspace{3.5em}+-- \textit{DepositMint} \newline
\textbar{} \newline
\textbar{}\hspace{0.5em}+-- \textit{\textbf{Economic Model}} \newline
\textbar{}\hspace{1.5em}+-- \textit{AssetBacking} \newline
\textbar{}\hspace{2.5em}+-- \textit{DeterministicMinting} \newline
\textbar{}\hspace{3.5em}+-- \textit{BalanceDeltaBased} \newline
\textbar{} \newline
\textbar{}\hspace{0.5em}+-- \textit{\textbf{Dependency}} \newline
\textbar{}\hspace{1.5em}+-- \textit{ERC20Token} \newline
\textbar{}\hspace{2.5em}+-- \textit{NonStandardERC20} \newline
\textbar{}\hspace{3.5em}+-- \textit{FeeOnTransfer}

&

\textit{\textbf{Failure Mode}} \newline
\textbar{} \newline
\textbar{}\hspace{0.5em}+-- \textit{\textbf{Failure Pattern}} \newline
\textbar{}\hspace{1.5em}+-- \textit{ERC20Incompatibility} \newline
\textbar{}\hspace{1.5em}+-- \textit{AccountingMismatch} \newline
\textbar{} \newline
\textbar{}\hspace{0.5em}+-- \textit{\textbf{Root Cause}} \newline
\textbar{}\hspace{1.5em}+-- \textit{FullTransferAssumption} \newline
\textbar{} \newline
\textbar{}\hspace{0.5em}+-- \textit{\textbf{Invariant Violation}} \newline
\textbar{}\hspace{1.5em}+-- \textit{CollateralConsistency} \newline
\textbar{}\hspace{1.5em}+-- \textit{Redeemability} \newline
\textbar{} \newline
\textbar{}\hspace{0.5em}+-- \textit{\textbf{Impact}} \newline
\textbar{}\hspace{1.5em}+-- \textit{OverMinting} \newline
\textbar{}\hspace{1.5em}+-- \textit{DoS}

&

\textit{\textbf{Exploit Primitive}} \newline
\textbar{} \newline
\textbar{}\hspace{0.5em}+-- \textit{\textbf{Setup}} \newline
\textbar{}\hspace{1.5em}+-- \textit{TokenDeployment} \newline
\textbar{}\hspace{1.5em}+-- \textit{TokenRegistration} \newline
\textbar{} \newline
\textbar{}\hspace{0.5em}+-- \textit{\textbf{Exploitation}} \newline
\textbar{}\hspace{1.5em}+-- \textit{RepeatedDeposits} \newline
\textbar{} \newline
\textbar{}\hspace{0.5em}+-- \textit{\textbf{Arbitrage and Exit}} \newline
\textbar{}\hspace{1.5em}+-- \textit{RepeatedWithdrawals} \\

\hline
\bottomrule
\end{tabular}

\end{table*}
\renewcommand{\arraystretch}{1}

\textbf{Case study: over-minting vulnerability causing portal insolvency}.
We analyze a 0-day vulnerability in the \texttt{MergeTokenPortal.sol} contract from the zkLink MergeToken project.
As shown in Listing~\ref{lst:case_study}, the contract functions as a portal, enabling asset conversion through a deposit–withdraw mechanism.
Users deposit a source token to receive a synthetic token, and can later burn the synthetic token to redeem the underlying asset. 
However, this design is incompatible with fee-on-transfer tokens, which deduct a fee on each transfer.
Because the \texttt{deposit} function mints synthetic tokens according to the \texttt{\_amount} rather than the actual amount received (Line 5), each deposit increases the synthetic token supply beyond the underlying collateral.
This discrepancy accumulates over time, ultimately resulting in insufficient reserves, failed withdrawals, and permanent losses for merge token holders.

Identifying this vulnerability is non-trivial for LLM-based analysis because it results from a compound semantic mismatch among external token behavior, fee-on-transfer mechanics, and internal collateral accounting, rather than from a naive local implementation flaw.
When reasoning is confined to the local \texttt{MergeTokenPortal}, the model lacks the necessary context to infer that an external token’s economic behavior can invalidate the portal’s internal accounting assumptions.

\setlength{\textfloatsep}{6pt}

\begin{lstlisting}[language = Solidity, caption={Simplified code snippet of the over-minting vulnerability in \texttt{MergeTokenPortal.sol}.},label=lst 1, float =t!, label = lst:case_study]
function deposit(address _sourceToken, ...) external {
    // ...
    IERC20Upgradeable(_sourceToken).safeTransferFrom(msg.sender, address(this), _amount);
    address mergeToken = tokenInfo.mergeToken;
    IERC20MergeToken(mergeToken).mint(_receiver, _amount);
    // ...}
function withdraw(address _sourceToken, ...) external {
    // ...
    require(tokenInfo.balance >= _amount, "Source Token balance is not enough");
    unchecked {tokenInfo.balance -= _amount;}
    address mergeToken = tokenInfo.mergeToken;
    IERC20MergeToken(mergeToken).burn(msg.sender, _amount);
    IERC20Upgradeable(_sourceToken).safeTransfer(_receiver, _amount);
    // ...}
\end{lstlisting}

Table~\ref{tab:case} illustrates the key nodes retrieved by \sys from the hierarchical knowledge graph during its analysis of the \texttt{MergeTokenPortal} contract, highlighting how structured knowledge supports multi-hop, cross-domain reasoning.
\sys first classifies the contract under the \textit{TokenWrapping} protocol category and further identifies it as a \textit{Portal} implementing a \textit{DepositMint} pattern.
By analyzing the \texttt{deposit} and \texttt{withdraw} functions, \sys infers that the contract adopts a balance-delta-based deterministic minting model with a strict 1:1 collateral backing assumption.
Through dependency-level retrieval, \sys identifies the involvement of a \textit{FeeOnTransfer} token, whose economic model imposes a fee on each transfer.
This behavior violates the portal’s implicit full-transfer assumption, revealing a semantic inconsistency between the contract’s internal accounting model and its external token dependency.
Guided by this semantic conflict, \sys reasons across the contract semantics layers to identify the contract’s failure pattern, characterized by an \textit{AccountingMismatch} that leads to systematic \textit{OverMinting} and violation of \textit{CollateralConsistency}, ultimately breaking \textit{Redeemability}.
Building on this identified failure mode, \sys links it to the corresponding exploit primitives in the HKG to generate a plausible PoC.
The PoC was validated through local execution, confirming the correctness of the inferred vulnerability, which was later acknowledged by the developers and fixed in a subsequent revision.

\begin{center}
  \small
  \fcolorbox{black}{gray!10}{\parbox{.97\linewidth}{\leftskip=0.6em \rightskip=0.6em \textbf{Answer to RQ4: } \sys demonstrates its real-world impact by detecting 16 previously unknown vulnerabilities, resulting in \$2,900 USD awards and securing protocols managing over \$70.6M USD.}}
\end{center}

\section{Discussion}

\subsection{Data Leakage and Generalization Capability}

The risk of data leakage is a threat to validity of experimental results in LLM-based systems.
We address this from three complementary angles.

\textbf{HKG construction data isolation.}
The long-term memory of \sys is constructed from data entirely disjoint from the evaluation datasets and stores only abstracted, reusable knowledge rather than instance-level code.

\textbf{Post-training cutoff cases.}
Table~\ref{tab:postdate} lists four cases of D2 whose public disclosure post-dates GPT-5's knowledge cutoff (May 2024), meaning the model cannot have seen their exploit details during pretraining. 
\sys successfully generates valid PoCs for all four cases with HKG guidance, while the no-HKG baseline produces none, directly ruling out memorization as the driver for these results.

\begin{table}[t]
\caption{Cases post-dating GPT-5 knowledge cutoff (May 2024).
\setlength{\tabcolsep}{5pt}
\ding{51}/\ding{55} denotes exploit success with/without HKG.}
\label{tab:postdate}
\centering
\scriptsize
\begin{tabular}{lcccc}
\toprule
\textbf{Case} & \textbf{Disclosure}
  & \textbf{Lag (mo.)}
  & \textbf{w/ HKG} & \textbf{w/o HKG} \\
\midrule
wifcoin    & Jun 2024 & +1  & \ding{51} & \ding{55} \\
pledge     & Dec 2024 & +7  & \ding{51} & \ding{55} \\
drlvaultv3 & Nov 2025 & +18 & \ding{51} & \ding{55} \\
yeth       & Dec 2025 & +19 & \ding{51} & \ding{55} \\
\bottomrule
\end{tabular}
\end{table}

\textbf{Memorization test.}
To further distinguish memorization from generalization, our ablation~\ref{sec_ab} already provides this test: the no-HKG baseline, which reduces \sys to direct LLM reasoning over raw contract code, produces zero successful exploits across all 88 cases (Table~\ref{tab:Ablation}).

\subsection{Reliability of Memory Evolution}
The effectiveness of \sys’s evolving memory is influenced by the quality of the underlying domain knowledge and vulnerability intelligence, which constitutes a potential threat to validity.
To mitigate this issue, we adopt a selection mechanism in Section~\ref{sec:HKG}, which requires candidate intelligence to satisfy at least two of three predefined criteria, thereby filtering out noisy or incomplete inputs.
Naturally, the system exhibits its strongest performance when provided with "gold-standard" intelligence that meets all three criteria.

\section{Related Work}

\textbf{LLM-based smart contract auditing}
leverages models’ understanding of code to detect vulnerabilities~\cite{10.1145/3597503.3639117,wu2024advscanner,yu2025sael,zhang2025acf,wei2025advanced}. 
\textsc{GPTScan}~\cite{10.1145/3597503.3639117} is the first to detect logical vulnerabilities in contracts using LLMs.
\textsc{iAudit}~\cite{10.1109/ICSE55347.2025.00027} combines fine-tuning and LLM-based agents for intuitive auditing with explanations, while \textsc{Smart-LLaMA-DPO}~\cite{yu2025smart} introduces preference-based optimization to improve vulnerability detection.
However, these approaches still face a semantic gap to bridge vulnerability identification and exploitable PoC generation.
\sys address this via a structured hierarchical knowledge graph and agentic memory, enabling reliable vulnerability detection and PoC synthesis.

\textbf{Smart contract fuzzing}
constructs transaction sequences to detect vulnerabilities during execution~\cite{10.1145/3597503.3639152}. 
Early approaches~\cite{Rethinking,lin2025automatic,lin2025promfuzz,rodler2023ef,10.1109/TDSC.2022.3182373} use control and data flow patterns as oracles, but are limited by predefined templates. 
Recent methods~\cite{Midas,VERITE} adopt profit-driven oracles to detect profitable vulnerabilities, serving partially as PoC generators. 
However, these fuzzers rely on hard-coded heuristics, limiting coverage and scalability, and struggle with complex contextual reasoning. 
\sys leverages domain knowledge-guided LLMs with verification process and an evolving memory to generate reliable PoCs and handle a broader range of contracts and vulnerabilities.

\textbf{Automated exploit generation} has been extensively studied in binary and web security, with systems such as AEG~\cite{avgerinos2011aeg} and Q~\cite{schwartz2011q} combining symbolic execution with vulnerability signatures to automatically produce working exploits from detected flaws.                                             
These works share the core challenge that motivates \sys: bridging the gap between a detected vulnerability and an executable attack artifact. 
However, DeFi exploits differ fundamentally: they involve inter-contract economic state manipulation, profitability requirements, and protocol-specific interaction semantics that have no direct analogue in traditional binary or web exploitation.

\section{Conclusion}
We present \sys, a knowledge-driven agentic system for end-to-end smart contract vulnerability detection and PoC generation. 
By organizing vulnerability intelligence into a hierarchical knowledge graph and leveraging self-evolving agentic memory with two-step verification, \sys achieves robust reasoning and exploit synthesis. 
Experiments show that \sys outperforms SOTA vulnerability scanners, fuzzers, and PoC generators in both vulnerability identification and exploit success rate. 
\sys also discovered 16 0-day vulnerabilities across five real-world projects.

\section*{Ethics Considerations}

\textbf{Research Scope and Ethical Boundaries.} 
This research strictly adheres to ethical standards for security and AI system evaluation. 
\sys is designed and presented solely for the purpose of analyzing and improving the robustness of DeFi protocol security. 
Our work aims to reveal the inherent limitations of current smart contract auditing practices and demonstrate the feasibility of automated exploit synthesis as a means to accelerate vulnerability validation. 
All experiments were conducted in controlled, locally hosted environments built upon Foundry's fork simulation framework.
No live, online, or third-party blockchain systems were accessed, attacked, tested, or influenced at any stage of this research. 
All exploit scripts execute against a locally forked blockchain state and do not submit any transactions to mainnet or any testnet.

\textbf{Zero-day Disclosure and Responsible Use.} 
The 0-day vulnerabilities discovered by \sys during our evaluation were responsibly disclosed to the respective project developers prior to public reporting. 
We reported all findings through official channels, including direct developer contact and the Secure3 bug bounty platform, and withheld technical details until developers had sufficient time to acknowledge and remediate the issues. 
To date, 16 vulnerabilities have been confirmed or patched by the respective teams. 
Only the high-level descriptions of vulnerability categories are included in the paper; no functional exploit code targeting any real-world deployed contract is distributed.

\textbf{Dual-use Considerations.} 
We acknowledge that automated exploit generation tools carry inherent dual-use risks. 
To mitigate potential misuse, the full system is intended for release to vetted security researchers and auditors only, with usage restricted to contracts for which the user holds authorization. 
We encourage the community to adopt \sys as a standard component in pre-deployment security audits, thereby strengthening the defensive posture of DeFi protocols before adversarial exploitation occurs.

\bibliographystyle{IEEEtran}
\bibliography{sample-base}

\appendices

\section{Walk-through of Two-stage Validation}
\label{app:two-stage-walkthrough}

This appendix illustrates how \sys validates an exploit candidate
using a real-world signature-bypass minting vulnerability. The
vulnerable token contract exposes a privileged \texttt{mint} function
that is intended to be protected by a multi-signature authorization
check. The exploit candidate bypasses this check using empty signature
arrays, mints \(10^{12}\) BEGO tokens to the attacker, and then
liquidates the minted tokens through the BEGO/WBNB AMM pair to realize
profit.

\subsection{Victim Contract Semantics}

Listing~\ref{lst:bego-vuln} shows the simplified vulnerable logic. The
contract provides a \texttt{mint} function that takes a mint amount, a
replay-protection identifier, a receiver address, and three arrays
representing ECDSA signature components. The intended design is that
\texttt{mint} should only proceed when the submitted signatures are
produced by authorized signers. To prevent replay, the function records
used identifiers in \texttt{txHashes}.

However, the authorization logic is flawed. The modifier
\texttt{isSigned} first checks only whether the three signature arrays
have equal lengths. It then allocates an array of recovered signers with
length \texttt{\_r.length}, fills this array by iterating over the
submitted signatures, and finally calls \texttt{isSigners} to verify
whether all recovered addresses are authorized signers. When the attacker
submits empty arrays, the length check succeeds because all three arrays
have length zero. The recovery loop is skipped, and \texttt{isSigners}
receives an empty signer array. Since \texttt{isSigners} only rejects
explicitly invalid signers inside the loop, the loop is also skipped and
the function returns \texttt{true}. As a result, the signature check is
bypassed and the privileged minting sink becomes reachable without any
valid signature.

\begin{lstlisting}[language=Solidity,caption={Simplified signature-bypass minting vulnerability.},float=t!,label={lst:bego-vuln}]
function mint(
    uint256 _amount,
    string memory _txHash,
    address _receiver,
    bytes32[] memory _r,
    bytes32[] memory _s,
    uint8[] memory _v
) isSigned(_txHash, _amount, _r, _s, _v)
  external returns (bool)
{
    require(!txHashes[_txHash], "tx-hash-used");
    txHashes[_txHash] = true;

    _mint(_receiver, _amount);
    return true;
}

modifier isSigned(
    string memory _txHash,
    uint256 _amount,
    bytes32[] memory _r,
    bytes32[] memory _s,
    uint8[] memory _v
) {
    require(checkSignParams(_r, _s, _v), "bad-sign-params");

    bytes32 _hash =
        keccak256(abi.encodePacked(bsc, msg.sender, _txHash, _amount));

    address[] memory _signers = new address[](_r.length);

    for (uint8 i = 0; i < _r.length; i++) {
        _signers[i] = ecrecover(_hash, _v[i], _r[i], _s[i]);
    }

    require(isSigners(_signers), "bad-signers");
    _;
}

function checkSignParams(
    bytes32[] memory _r,
    bytes32[] memory _s,
    uint8[] memory _v
) internal pure returns (bool) {
    return _r.length == _s.length && _s.length == _v.length;
}

function isSigners(address[] memory _signers)
    public view returns (bool)
{
    for (uint8 i = 0; i < _signers.length; i++) {
        if (!_containsSigner(_signers[i])) {
            return false;
        }
    }
    return true;
}
\end{lstlisting}

\subsection{Candidate Exploit Plan}

For readability, we use \(\mathrm{swapBEGOToWBNB}\) to denote the
PancakeSwap V2 router call that liquidates the minted BEGO tokens into
WBNB. Given the above semantics, \sys constructs the following
high-level exploit plan:
{\small
\[
\begin{aligned}
P &= \langle t_1,t_2,t_3\rangle, \\
t_1 &=
\langle
C_{\mathrm{BEGO}},
\mathrm{mint},
\sigma_{\mathrm{mint}},
K_{\mathrm{mint}}
\rangle, \\
t_2 &=
\langle
C_{\mathrm{BEGO}},
\mathrm{approve},
\sigma_{\mathrm{approve}},
K_{\mathrm{allowance}}
\rangle, \\
t_3 &=
\langle
C_{\mathrm{Router}},
\mathrm{swapBEGOToWBNB},
\sigma_{\mathrm{swap}},
K_{\mathrm{profit}}
\rangle.
\end{aligned}
\]
}

The first interaction targets the vulnerable token contract and attempts
to reach the privileged minting operation. The mint sink and its symbolic
input are:
{\small
\[
\begin{aligned}
&K_{\mathrm{mint}}
= \mathrm{\_mint}(\mathit{attacker},M_{\mathrm{raw}}), \\
&\sigma_{\mathrm{mint}}
= (M_{\mathrm{raw}},h,\mathit{attacker},R,S,V).
\end{aligned}
\]
}
Here, \(h\) is an attacker-chosen replay-protection identifier that has
not been used before, and \(R,S,V\) are attacker-controlled signature
arrays. The candidate assigns:
{\small
\[
\begin{aligned}
&M = 10^{12}\ \mathrm{BEGO}, \\
&M_{\mathrm{raw}} = 10^{12}\times 10^{18}, \\
&|R| = |S| = |V| = 0.
\end{aligned}
\]
}
We use \(M\) for the human-readable token amount and
\(M_{\mathrm{raw}}\) for the 18-decimal ERC-20 base-unit amount.
After minting, the attacker approves the PancakeSwap V2 router to spend
the minted BEGO tokens and swaps all received BEGO for WBNB. 
The resulting exploit path is:
{\small
\[
\begin{aligned}
&\mathrm{BEGO.mint}(M_{\mathrm{raw}},h,\mathit{attacker},[],[],[]) \\
&\rightarrow \mathrm{isSigned}
\rightarrow \mathrm{checkSignParams}
\rightarrow \mathrm{skip\mbox{-}ecrecover\mbox{-}loop} \\
&\rightarrow \mathrm{isSigners}([])
\rightarrow \mathrm{\_mint}(\mathit{attacker},M_{\mathrm{raw}}) \\
&\rightarrow \mathrm{BEGO.approve}(\mathrm{Router},\infty)
\rightarrow \mathrm{\scalebox{0.97}{Router.swapBEGOToWBNB}}.
\end{aligned}
\]
}

\begin{table*}[t]
\centering
\caption{Step-wise asset-state simulation for the BEGO signature-bypass minting exploit.}
\label{tab:bego-asset-sim}
\begin{tabular}{lcccc}
\toprule
Step & Operation & Attacker BEGO & Attacker WBNB & AMM state \(\Omega\) \\
\midrule
\(S_0\) & Initial state
& \(0\) & \(0\)
& \((R_{\mathrm{BEGO}},R_{\mathrm{WBNB}})\) \\

\(S_1\) & Prepare empty signatures
& \(0\) & \(0\)
& \((R_{\mathrm{BEGO}},R_{\mathrm{WBNB}})\) \\

\(S_2\) & Pass \texttt{isSigned}
& \(0\) & \(0\)
& \((R_{\mathrm{BEGO}},R_{\mathrm{WBNB}})\) \\

\(S_3\) & Execute mint sink
& \(10^{12}\) & \(0\)
& \((R_{\mathrm{BEGO}},R_{\mathrm{WBNB}})\) \\

\(S_4\) & Approve router
& \(10^{12}\) & \(0\)
& \((R_{\mathrm{BEGO}},R_{\mathrm{WBNB}})\) \\

\(S_5\) & Transfer BEGO to pair
& \(\approx 0\) & \(0\)
& \((R_{\mathrm{BEGO}}+x,R_{\mathrm{WBNB}})\) \\

\(S_6\) & Pair outputs WBNB
& \(\approx 0\) & \(out_{\mathrm{WBNB}}\)
& \((R_{\mathrm{BEGO}}+x,R_{\mathrm{WBNB}}-out_{\mathrm{WBNB}})\) \\

\(S_f\) & Final state
& \(\approx 0\) & \(out_{\mathrm{WBNB}}\)
& \(\Omega_f\) \\
\bottomrule
\end{tabular}
\end{table*}

\subsection{Stage 1: Exploit-path Reachability}

The first validation stage checks whether the privileged minting sink is
logically reachable from the public entry function \texttt{mint}. For the
candidate above, \sys performs semantic traversal from \texttt{mint}
through the \texttt{isSigned} modifier and collects the branch predicates
that guard \(\mathrm{\_mint}\). The extracted path is:
{\small
\[
\begin{aligned}
\pi =
\langle
&\mathrm{mint},
\mathrm{isSigned},
\mathrm{checkSignParams}, \\
&\mathrm{skip\mbox{-}ecrecover\mbox{-}loop},
\mathrm{isSigners},
\mathrm{\_mint}
\rangle .
\end{aligned}
\]
}
The path predicates collected from the contract are:
{\small
\[
\begin{aligned}
\phi_1 &: |R| = |S|, \\
\phi_2 &: |S| = |V|, \\
\phi_3 &: \neg \mathrm{txHashes}[h].
\end{aligned}
\]
}
The candidate-controlled assignments are:
{\small
\[
\begin{aligned}
&|R| = |S| = |V| = 0, \\
&\mathit{receiver} = \mathit{attacker}, \\
&M_{\mathrm{raw}} = 10^{12} \times 10^{18}.
\end{aligned}
\]
}
The signature recovery loop is guarded by \(0\leq i<|R|\). Since
\(|R|=0\), the loop has no feasible iteration:
{\small
\[
\forall i,\; \neg(0\leq i<0).
\]
}
Therefore, no call to \texttt{ecrecover} is required, and the recovered
signer array remains empty. Since the rejection condition in
\texttt{isSigners} is only evaluated inside the loop, the empty signer
array is accepted:
{\small
\[
\mathit{signers}=[ ]
\Longrightarrow
\mathrm{isSigners}(\mathit{signers})=\mathsf{true}.
\]
}
The SMT query used for this path consists of the collected predicates
instantiated with the empty-array assignment:
{\small
\[
\Phi =
\phi_1 \land \phi_2 \land \phi_3
\land (|R|=0) \land (|S|=0) \land (|V|=0).
\]
}
A satisfying assignment is:
{\small
\[
|R|=|S|=|V|=0,\quad
M_{\mathrm{raw}}=10^{12}\times 10^{18},
\]
\[
\mathit{receiver}=\mathit{attacker},\quad
\mathrm{txHashes}[h]=\mathsf{false}.
\]
}
Thus, the result is \(\textsc{SAT}(\Phi)=\mathsf{true}\), which confirms
that the exploit-critical sink in \(t_1\) is reachable:
\(\mathrm{\_mint}(\mathit{attacker},M_{\mathrm{raw}})\).
The remaining approval and swap interactions are then handled in the
asset-level profitability simulation.

\subsection{Stage 2: Profit Realizability}

The second validation stage checks whether the reachable minting sink can
be converted into positive asset-level profit. \sys tracks the
attacker's BEGO and WBNB balances, as well as the BEGO/WBNB AMM state,
throughout the candidate exploit. Unlike flash-loan-based candidates,
this exploit does not require upfront capital in the abstract asset
model; the input BEGO is created by the vulnerable mint operation.

We model the asset state as \(S=\langle B,\Omega\rangle\), where \(B\)
records account-level token balances and \(\Omega\) records the
BEGO/WBNB AMM state. The initial state is:
{\small
\[
\begin{aligned}
&\Omega_0 = (R_{\mathrm{BEGO}},R_{\mathrm{WBNB}}), \\
&B_0(\mathit{attacker},\mathrm{BEGO}) = 0, \\
&B_0(\mathit{attacker},\mathrm{WBNB}) = 0.
\end{aligned}
\]
}
After Stage~1 proves that empty signature arrays satisfy the
authorization modifier, the attacker calls
\(\mathrm{mint}(M_{\mathrm{raw}},h,\mathit{attacker},[],[],[])\). This
increases the attacker's BEGO balance by \(M\):
{\small
\[
B_3(\mathit{attacker},\mathrm{BEGO})=10^{12},
\quad
B_3(\mathit{attacker},\mathrm{WBNB})=0.
\]
}
The subsequent approval only updates allowance and does not change token
balances, so \(B_4=B_3\).
During liquidation, the router transfers the minted BEGO from the
attacker to the BEGO/WBNB pair. Let \(x\) denote the actual amount of
BEGO credited to the pair. For a standard ERC-20 transfer,
\(x=10^{12}\ \mathrm{BEGO}\); if the token applies transfer-side
deductions, \(x\) denotes the post-transfer amount received by the pair.
The AMM output is:
{\small
\[
out_{\mathrm{WBNB}}
=
\frac{\gamma x R_{\mathrm{WBNB}}}
     {R_{\mathrm{BEGO}}+\gamma x},
\]
}
where \(\gamma\) is the fee-adjusted input ratio. After the swap:
{\small
\[
B_f(\mathit{attacker},\mathrm{BEGO})\approx 0,
\quad
B_f(\mathit{attacker},\mathrm{WBNB})=out_{\mathrm{WBNB}}.
\]
}
Using WBNB as the numeraire and ignoring gas in the abstract simulation,
the asset-level wealth change is:
{\small
\[
\Delta W
=
\mathrm{Val}(B_f,\Omega_f)-\mathrm{Val}(B_0,\Omega_0)
=
out_{\mathrm{WBNB}}.
\]
}
When \(out_{\mathrm{WBNB}}>0\), the candidate is economically viable and
passes the second validation stage.
Table~\ref{tab:bego-asset-sim} summarizes the step-wise asset-state
simulation. The simulation includes the authorization steps for
continuity; they do not change asset balances.
The concrete asset transition is:
{\small
\[
(0\ \mathrm{BEGO},0\ \mathrm{WBNB})
\rightarrow
(\approx 0\ \mathrm{BEGO},out_{\mathrm{WBNB}}\ \mathrm{WBNB}).
\]
}
Since \(out_{\mathrm{WBNB}}>0\), the exploit realizes positive profit
and is forwarded for concrete Foundry execution.

This walk-through demonstrates how the two-stage validation framework
separates exploit feasibility into two complementary checks. The first
stage verifies that the security-critical sink is reachable under
attacker-controlled inputs, while the second stage verifies that reaching
this sink yields a positive asset-level outcome.

\begin{table}[t!]
\caption{Bootstrapped HKG node statistics and cross-layer edge statistics.}
\label{tab:hkg_stats}
\centering
\small
\begin{tabular}{lcc}
\toprule
\textbf{Layer} & \textbf{Nodes} & \textbf{Cross-layer Edges} \\
\midrule
Contract Semantics & 144 & CS$\to$FM: 165 \\
Failure Mode & 107 & FM$\to$EP: 132 \\
Exploit Primitive & 523 & -- \\
\midrule
\textbf{Total} & \textbf{774} & \textbf{297} \\
\bottomrule
\end{tabular}
\end{table}

\section{HKG Ontology Schema Details}
\label{app:hkg}

Table~\ref{tab:ontology} provides the complete ontology schema
of the Hierarchical Knowledge Graph, including all node types,
their semantic roles, intra-layer edge types, and cross-layer
linking constraints. The schema is fixed prior to HKG
construction and is not modified during knowledge abstraction
or fusion; see Section~\ref{sec:HKG} for the formal definition.

\noindent\textbf{Node type assignment.}
During knowledge abstraction, the LLM assigns each extracted
element a node type from the predefined schema via
chain-of-thought prompting. A node receives type $T$ only if
its semantic description matches the role definition of $T$ and
its incident edges are admissible under the schema constraints.
Cases where two candidate types score within a confidence margin
of 0.1 are flagged as ambiguous and resolved during fusion by
adopting the most consistent typing across similar cases.

\noindent\textbf{HKG statistics.}
As shown in Table~\ref{tab:hkg_stats}, the bootstrapped HKG contains 774 nodes across three layers: 144 contract semantics nodes, 107 failure mode nodes, and 523 exploit primitive nodes. 
It also includes 297 cross-layer edges, with 165 CS$\to$FM edges and 132 FM$\to$EP edges.
This moderate graph density reflects the design principle of encoding causal relevance rather than exhaustive linkage.

\begin{table*}[t]
\caption{HKG ontology schema: node types, semantic roles,
and cross-layer edge types.}
\label{tab:ontology}
\centering
\small
\setlength{\tabcolsep}{7pt}
\begin{tabular}{p{2.2cm}p{2.4cm}p{11.0cm}}
\toprule
\textbf{Layer} & \textbf{Node Type} & \textbf{Semantic Role} \\
\midrule
\multirow{4}{*}{\parbox{2.2cm}{Contract\\Semantics\\(CS)}}
  & Protocol
    & Top-level protocol category; serves as semantic anchor linked to all other CS nodes via \texttt{enforces}, \texttt{adopts}, \texttt{depends\_on}. \\[2pt]
  & Access Control
    & Permission structure and role enforcement logic of the contract (e.g., owner-only guards, proxy admin). \\[2pt]
  & Economic Model
    & Fee structures, tokenomics, reserve accounting, and incentive mechanisms (e.g., fee-on-transfer, rebase). \\[2pt]
  & Dependency
    & External contract and non-standard token dependencies that affect protocol behavior. \\
\midrule
\multirow{5}{*}{\parbox{2.2cm}{Failure\\Mode\\(FM)}}
  & Failure Pattern
    & Named vulnerability pattern (e.g., price manipulation, accounting mismatch); causally linked to root cause via \texttt{caused\_by}. \\[2pt]
  & Condition
    & Preconditions in the execution environment required to trigger the failure (e.g., flash loan availability). \\[2pt]
  & Root Cause
    & Underlying implementation or design flaw that enables the vulnerability (e.g., full-transfer assumption). \\[2pt]
  & Impact
    & Observable consequence of a successful exploit: fund drainage, over-minting, DoS, unauthorized withdrawal. \\[2pt]
  & Invariant Violation
    & Protocol-level invariant broken when the failure manifests (e.g., collateral consistency, reserve balance). \\
\midrule
\multirow{4}{*}{\parbox{2.2cm}{Exploit\\Primitive\\(EP)}}
  & Exploit Behavior
    & Generic, reusable exploit action template covering preparation, exploitation, and arbitrage phases. \\[2pt]
  & Specific Plan
    & Case-specific exploit step sequence grounded in a concrete DeFi incident; composed from exploit behaviors. \\[2pt]
  & PoC Framework
    & High-level Foundry test scaffold that assembles specific plans into a complete executable PoC. \\[2pt]
  & Few-shot Example
    & Concrete Solidity snippet grounding abstract exploit behaviors in real attack code for LLM reasoning. \\
\midrule
\multicolumn{3}{l}{\textbf{Cross-layer Edge Types}} \\
\midrule
\multicolumn{2}{l}{\texttt{related\_exploit}: CS $\to$ FM}
  & Links a CS node to failure patterns it is implicated in. \\[2pt]
\multicolumn{2}{l}{\texttt{caused\_by}: FP $\to$ RC}
  & Failure pattern is caused by a specific root cause. \\[2pt]
\multicolumn{2}{l}{\texttt{leads\_to}: RC $\to$ Inv}
  & Root cause leads to a protocol invariant violation. \\[2pt]
\multicolumn{2}{l}{\texttt{related\_exploit}: FM $\to$ EP}
  & Root cause pattern linked to corresponding exploit primitives. \\[2pt]
\multicolumn{2}{l}{\texttt{start\_at}: PocFrame $\to$ SpecificPlan}
  & PoC framework references specific plans as composable phases. \\
\bottomrule
\end{tabular}
\end{table*}

\end{document}